# Anisotropy of the chiral, semiconducting phase $LaRhC_2$: a handedness resolved study

Volodymyr Levytskyi[1,*], Ulrich Burkhardt[2,*], Markus König[2], Christoph Hennig[3,4], Eteri Svanidze[2], Yuri Grin[2], and Roman Gumeniuk[1,*]

[1] Institut für Experimentelle Physik, TU Bergakademie Freiberg, Leipziger Str. 23, 09599 Freiberg, Germany

[2] Max-Planck-Institut für Chemische Physik fester Stoffe, Nöthnitzer Str. 40, 01187 Dresden, Germany

[3] Helmholtz-Zentrum Dresden-Rossendorf, Institute of Resource Ecology, Bautzner Landstr. 400, 01314 Dresden, Germany

[4] Rossendorf Beamline (BM20-CRG), European Synchrotron Radiation Facility, 71 Avenue des Martyrs, 38043 Grenoble, France

*Corresponding authors' E-mails: volodymyr.levytskyi@physik.tu-freiberg.de
ulrich.burkhardt@cpfs.mpg.de
roman.gumeniuk@physik.tu-freiberg.de

## Abstract

Chirality in quantum materials is a topic of significant importance due to its profound effects on the electronic, magnetic, and optical properties of these systems. However, it is non-trivial to decouple the behavior of two enantiomorphs within the same material – perhaps explaining why the influence of chirality on electrical properties has remained largely unexplored. In this work, we examine the electrical conductivity, magnetoresistance, and thermal expansion coefficient of $LaRhC_2$ – a compound with a chiral crystal structure (tetragonal symmetry, space groups $P4_1$ or $P4_3$). The identification of a suitable monochiral domain was achieved via electron backscatter diffraction, which simultaneously determines crystallographic orientation and handedness. Monochiral specimens of $LaRhC_2$ enantiomorphs were further characterized by single-crystal X-ray diffraction, confirming the $P4_1$ and $P4_3$ symmetries. The analysis of electrical resistivity was made possible through the single-domain extraction of enantiopure specimens from a polycrystalline sample using focused ion beam techniques. We establish that $LaRhC_2$ is a semiconductor with band gaps of approximately 20 meV and 33 meV parallel and perpendicular to the fourfold screw axis of the crystal structure, respectively – consistent with band structure calculations. A significant anisotropy is also observed in the thermal expansion: $5.4 \times 10^{-6}$ $K^{-1}$ along the [100] and $2.8 \times 10^{-6}$ $K^{-1}$ along the [001] crystallographic directions of the tetragonal lattice. Moreover, angular-dependent longitudinal resistivity along the fourfold screw axis under magnetic field also exhibits pronounced anisotropy. This work clearly demonstrates that even in materials without topological order, chirality can influence electron transport and significantly impact electrical resistivity.

## Introduction

Materials with non-centrosymmetric crystal structures (NCCS) continue to be of particular interest in solid state chemistry and condensed matter physics, because properties like pyroelectricity, ferroelectric orderings, piezoelectric and/or optical activity[1,2] require the lack of an inversion center in the crystal structure. Recently, novel properties of topological materials have emerged – conductive surface states in insulators[3], unusual spin arrangements such as spin textures (skyrmions)[4], and magnetotransport phenomena including spin Hall effect and chiral anomaly[5]. For many of these systems, a complex relationship between the structural chirality and the resultant spins states have been proposed. As a result, NCCS-materials are now considered as a key to evolution of advanced technologies as nonlinear optics, quantum computing, and spintronics.[1,6,7,8] However, since a complete understanding of bulk vs. surface effects is still lacking, it is important to consider if any of the peculiar phenomena observed in topologically non-trivial materials can be found is systems for which the role of topology is negligible. In particular, understanding the effect of chirality on electrical properties of various materials have so far mostly been focused on topological materials with chiral space group (SG) symmetry crystal structures[1], with less work dedicated to topologically trivial systems, also showing interesting spin and transport properties[7,9,10]. For instance, the chiral metallic systems $M\mathrm{Rh_2B_2}$ ($M$ = Ta, Nb; SG $P3_1$) undergo a superconducting transition. However, this quantum state has only been studied in polycrystalline samples.[11] A more specific experiment, *viz*. de Haas–van Alphen experiment, on the single crystalline enantiomorphs of binary metallic compounds $M\mathrm{Si_2}$ ($M$ = Ta, Nb, V) revealed a split of the Fermi surface, thus reflecting non-centrosymmetric structure.[12] By examining the effects of structural chirality on electrical transport in topologically trivial non-magnetic semiconducting materials, this work contributes to understanding the broader role of symmetry-breaking in quantum transport beyond topological considerations.

In this sense, $LaRhC_2$ is a perfect candidate – it is a semiconductor with a trivial band gap without band crossing points close to the Fermi level. Recent calculations based on a minimal tight-binding model indicate that $LaRhC_2$ exhibits directionally dependent topological surface and edge states.[13] Furthermore, several closely related compounds $RMC_2$ ($R$ = rare earth element, $M$ = transition metal) have been identified topological Weyl or Dirac (for $GdRuC_2$) semimetals[14,15] with significant influence on the chiral and gyrotropic magnetic effects and strong dependence from the canting angle between external magnetic field and the crystal structure. The physical properties on the bulk polycrystalline samples were reported by Hoffmann *et al.*,[16] and revisited recently by Liu and co-authors[17]. From these studies, $LaRhC_2$ appears to be semiconducting and diamagnetic. Interestingly, the electronic band structure reports on the $LaRhC_2$ are contradictory: it is shown to be a semiconductor with a narrow energy gap of 84 meV in[14], whereas Liu *et al.*[17] accounting for spin-orbit coupling found a semimetallic state.

Chirality-dependent properties are difficult to verify, as they are often small, direction-dependent and can be reduced or even completely canceled out by superposition. Progress in surface handedness analysis allowed to resolve and extract the individual enantiomorphs of such chiral bulk polycrystalline intermetallic materials as Weyl semimetal CoSi (SG $P2_13$)[18], magnetic MnSi[19], charge-ordered $CsV_3Sb_5$[20], or even simple metals, like β-Mn[21]. Manufacturing a crystallographically oriented microdevices of a given handedness and subsequent investigation of their electrical transport properties is fairly demanding. However, it requires specific and advanced techniques in chemistry and physics. Our investigations were carried out on single-domain samples with different chirality, extracted from a polycrystalline material. Our previous advances in micro-scale isolation of selected phases within a multi-phase material have served as a stepping stone for the realization of this non-trivial task.[15,22,23] The measurements of $LaRhC_2$ were performed parallel and perpendicular to the tetragonal axis enabled the assessment of the impact of the structural anisotropy on the resultant properties. This work can clearly disentangle crystal and electronic structure as well as thermal expansion coefficients of right-handed (SG $P4_1$) and left-handed (SG $P4_3$) enantiomorphs of semiconducting $LaRhC_2$. This first example of anisotropically-resolved electrical conductivity and magneto-resistivity on two enantiomorphs with the same chemical composition give insight on the underlying role of chirality in solid-state materials with conventional topology.

## Experiment

### Sample preparation and PXRD analysis

Bulk polycrystalline samples with starting compositions close to $LaRhC_2$ [i.e., $La_{1\pm x\mp y}Rh_{1\mp x\mp z}$, $(0 \leq x, y, z \leq 0.12)$] and total mass of 0.5 g were arc-melted in argon-filled setup. A titanium metal button as getter was arc-melted first for additional purification the argon atmosphere. The weight losses after three times re-melting were usually below 1 wt. %. For the syntheses, lanthanum (ChemPur, 99.9 wt. %), rhodium (Degussa, 99.9 wt. %) metals and a graphite rod (Chempur 99.9995 wt. %) were used as starting components. To avoid additional contaminations, all handling and manipulations were performed inside an argon-filled glove box [MBraun, $p(O_2/H_2O) \leq 1$ ppm].

Homogenization of the alloys was performed at 1070 K and 1520 K. In first case, the samples were directly placed into silica ampoules, evacuated and then sealed. After annealing, the ampoules with samples inside were quenched in a cold water. In second case, the samples were placed into alumina crucibles, enclosed under argon in tantalum containers, and then additionally sealed into evacuated silica tubes and annealed without quenching for 7 days.

The as-cast and annealed polycrystalline samples were characterized by powder X-ray diffraction (PXRD) method (Huber G670 imagine plate Guinier camera equipped with Cu-anode tube and curved Ge-[111] monochromator ($K\alpha_1$-radiation, $\lambda = 1.54056$ Å, $3° < 2\theta < 100°$, $\Delta(2\theta) = 0.005°$). The phase analysis and crystallographic calculations were performed with STOE *WinXPOW* and *WinCSD* program packages[24,25] by comparing the theoretical profiles of binary and ternary phases (crystallographic information obtained from ICSD database)[26] with the experimental ones. Refinements of the unit cell parameters (UCPs) were carried out by least-squares full-profile fitting (*WinCSD* software) of the PXRD patterns measured with $LaB_6$ (NIST-660b) internal standard.

Synchrotron powder XRD data were collected at the Rossendorf Beamline BM20 (ESRF, Grenoble, France)[27] at different temperatures (100 K $\leq T \leq$ 295 K). The specimen was enclosed in an argon filled Lindemann glass capillary of diameter 0.3 mm. The XRD images were recorded by the Dectris Pilatus3 X 2M detector (radiation $\lambda = 0.61996$ Å) with a counting time of 40 s per frame and a cooling rate of 4 K $min^{-1}$. Diffractometer control and data collection was performed using *Pylatus* software[28]. The temperature was controlled by $N_2$-cryostream. The unit cell parameters (UCPs) vs. $T$ were refined in sequential mode by full profile fitting using the *FullProf* program package[29].

**SEM and EBSD-based chirality analyses**

Metallographic analysis of the polycrystalline material was performed by using optical microscopy (Zeiss Axioplan 2, bright-field and polarized light) as well as scanning electron microscopy (SEM; JEOL JSM 7800F). The electron backscatter diffraction (EBSD) patterns were recorded with the EBSD detector ($e^-$-Flash$^{HR}$ detector, Brukernano) attached to the SEM. The pattern evaluation was performed with the program *Quantax Esprit* (ver.3.2, Brukernano) to determine the grain orientation and inverse pole figures. The assignment of the handedness is realized by the evaluation with the EBSD pattern-matching approach, implemented in the software module *Mapsweeper* (software package *AZtecCrystal*, version 3.2 Oxford instruments). Here, the comparison between measured and simulated EBSD pattern is quantified by the calculation of the cross-correlation coefficient $r$ ($0 < r < 1$), with $r \approx 1$ for an almost complete match.[30] The key property for the assignment of the handedness is the consideration of the dynamical electron scattering processes in the EBSD pattern simulation. It leads to chirality-dependent contributions to the simulated EBSD pattern and to a different match with the experimental EBSD pattern, e.g. different cross-correlation coefficients $r_+$, $r_-$ for both enantiomorphs. For $LaRhC_2$ an additional test for the best match is necessary for patterns related by the rotation the crystal structure by 180° around [110] and its inverted crystal structure. According to our experience, the difference in the $r$-values should exceed 0.01 ($\Delta r = |r_+ - r_-| > 0.01$) for a reliable assignment of the handedness. Pre-conditions are high quality experimental EBSD patterns with reasonably good match with the simulated ones e.g., $r_+, r_- \geq 0.5$.[31,32] The simulated pattern with the best match with the measured ones is used to assign the handedness. The observed differences $\Delta r \approx 0.05$ for $LaRhC_2$ are large compared to other phases we have investigated [e.g. $\Delta r \approx 0.02$ was found for the chiral phase CoSi of the FeSi (B20) type of structure, space group $P2_13$][33].

**Monochiral single crystals and microdevices: preparation and electrical measurements**

Monochiral single-crystalline cubes and thin lamellae were cut from the polycrystalline material by using the focused ion beam (FIB) method (FEI Helios G4 PFIB machine, Xe-beam, current $I < 200$ nA, acceleration voltage $U = 30$ kV).

Single crystal X-ray diffraction (scXRD) experiments were performed on the FIB-cut cubes ($l \approx 30$ µm), glued with the NVH oil (Jena Bioscience) to the micro-grippers (MiTeGen). ScXRD diffraction images were recorded in rotational mode ($\omega$-scans) at temperatures $T$ = 293 K, 200 K, and 100 K on a STOE STADIVARI diffractometer (Mo$K\alpha$-radiation, $\lambda$ = 0.71073 Å) equipped with Dectris Pilatus300K detector and $N_2$-cryocooler. The reflection intensities, corrected by absorption (numerical/integration)[34] and scaled, were extracted by *X-AREA* software package[35]. Initial structural models were obtained by direct methods using *SIR*-2019[36], and refinement by *SHELX*-2018/3 program[37] (*WinGX* package[38]).

The thin lamellae were shaped into microdevices using the FIB method, with which the electrical resistivity in standard ($\rho_{xx}$) and Hall ($\rho_{xy}$) four-probe configuration can be measured both parallel and perpendicular ($\rho^{\parallel}$ and $\rho^{\perp}$) to the fourfold screw axis of the tetragonal crystal lattice of the $LaRhC_2$ phase. The *ac* electrical transport- (ETO) or *dc* resistivity- options of the DynaCool-12 (Quantum Design) device were used for the measurements in magnetic fields with 0 and 10 T and at temperature range of 1.8–300 K. The dependence of $\rho^{\parallel}_{xx}$ on the angle between fourfold screw axis and the magnetic field (10 T) was determined using the horizontal rotator. The amplitudes of the applied electrical currents did not exceed 10 μA.

**Electronic structure calculations**

The first-principles electronic structure calculations for $LaRhC_2$ have been performed within the local density approximation (LDA) of the density-functional theory (DFT) using the full-potential *FPLO* code (version 9.01)[39]. In the scalar relativistic calculation, the exchange-correlation potential by Perdew and Wang was used[40]. The *k*-mesh included 8000 points in the full Brillouin zone.

## Results

### From polycrystalline chiral $LaRhC_2$ to its single-crystalline enantiomorphs

$LaRhC_2$ was obtained in the samples of initial stoichiometric or slightly off-stoichiometric mixture of the elements. This also led to the crystallization of additional minority phases beside the target phase (see *Synthesis optimization* in **Supplementary Materials**). Practically, the absence of UCPs variation of the $LaRhC_2$, also in the off-stoichiometric multi-phases samples (within the provided standard errors, Fig. S1, S2), and their full agreement with the standardized ($\alpha$-$SiO_2$ as internal reference) ones for the sample reported by Hoffmann *et al.* [$a$ = 3.9697(5) Å, $c$ = 15.333(4) Å][16] strongly indicate that chemical composition of the $LaRhC_2$ compound remained unchanged.

The single-crystalline monochiral specimens of $LaRhC_2$ were extracted from the annealed sample of the initial La : Rh : C = 1 : 1 : 2 molar stoichiometry. SEM analysis of the microstructure of the cross-section indicated the $LaRh_2$ and $LaC_2$ were situated at the grain boundaries of $LaRhC_2$ majority phase. Due to the strong orientational contrast of different phases, it was possible to use polarized light optical microscopy of the polycrystalline material to pre-select regions with grains of several $\times 10^2$ µm in cross-section. The EBSD analysis revealed that the grain orientation was not textured and, surprisingly, many grains contained two or more domains of different chirality. The enantiomorph distribution appears to be balanced, with equal amounts of left- and right-handed domains. This makes identification of large (>100 µm) monochiral domains possible. Two of such domains were used for the FIB extraction of the cubic specimens for the scXRD experiments, as discussed in the next section. Other domains were selected for the preparation of the transport microdevices.

A schematic process for the microdevice preparation is shown in Fig. 1, with an example of $P4_3$-$LaRhC_2$. Only domains with an almost horizontal orientation of the fourfold screw axis (e.g. perpendicular to the surface normal) were used. Suitable grains were highlighted in the inverse pole figure Z (IPF-Z) map as a result of an acceptance range of 10° around the equatorial line, defined in the pole figure representation. In the initial step, two cuts parallel to the fourfold screw axis were made perpendicular to the surface of the cross-section. Subsequently, thin plates (lamellae) were lifted out from the polycrystalline material and fixed on a substrate. Chirality and crystallographic orientation of the lamella were analyzed by EBSD prior to the final shaping of the device. The final device contains two series-connected bars to measure the longitudinal electrical resistivity ($\rho_{xx}$) parallel ($\rho^{\parallel}$) and perpendicular ($\rho^{\perp}$) to the fourfold screw axis (i.e., [001] crystallographic direction), which means respective current directions $\vec{j} \parallel [001]$ and $\vec{j} \perp [001]$. Additional contacts were also made for the transverse electrical resistivity ($\rho_{xy}$) probes. The bars were cut to have virtually the same cross-section for a given microdevice. Depending on the final size of the lamella, the voltage measurement points ($V_{xx}/V_{xy}$) were separated by $l_{xx} \approx 10\ldots26$ µm for longitudinal and $l_{xy} \approx 2.8\ldots5.0$ µm for transverse measurements, respectively (Tables S2, S3). The cross-section areas varied between 4.4 and 16.3 µm$^2$. The misalignments between the current direction and the fourfold screw axis were smaller than 10° in the case of $\rho^{\parallel}$ measurements. For the case of $\rho^{\perp}$, the current directions were found to be almost parallel to [100] or [110] crystallographic directions. The dimensions (effective bar-shaped volumes used in the measurements) and the crystallographic orientations of the microdevices are listed in Tables S2, S3.

Interestingly, EBSD analysis of the grains with horizontal orientation of the screw axis showed that the interface between the domains of different handedness is oriented perpendicular to the fourfold screw axis. A detailed study on the twinning irregularities of $LaRhC_2$ are the subject of another study.

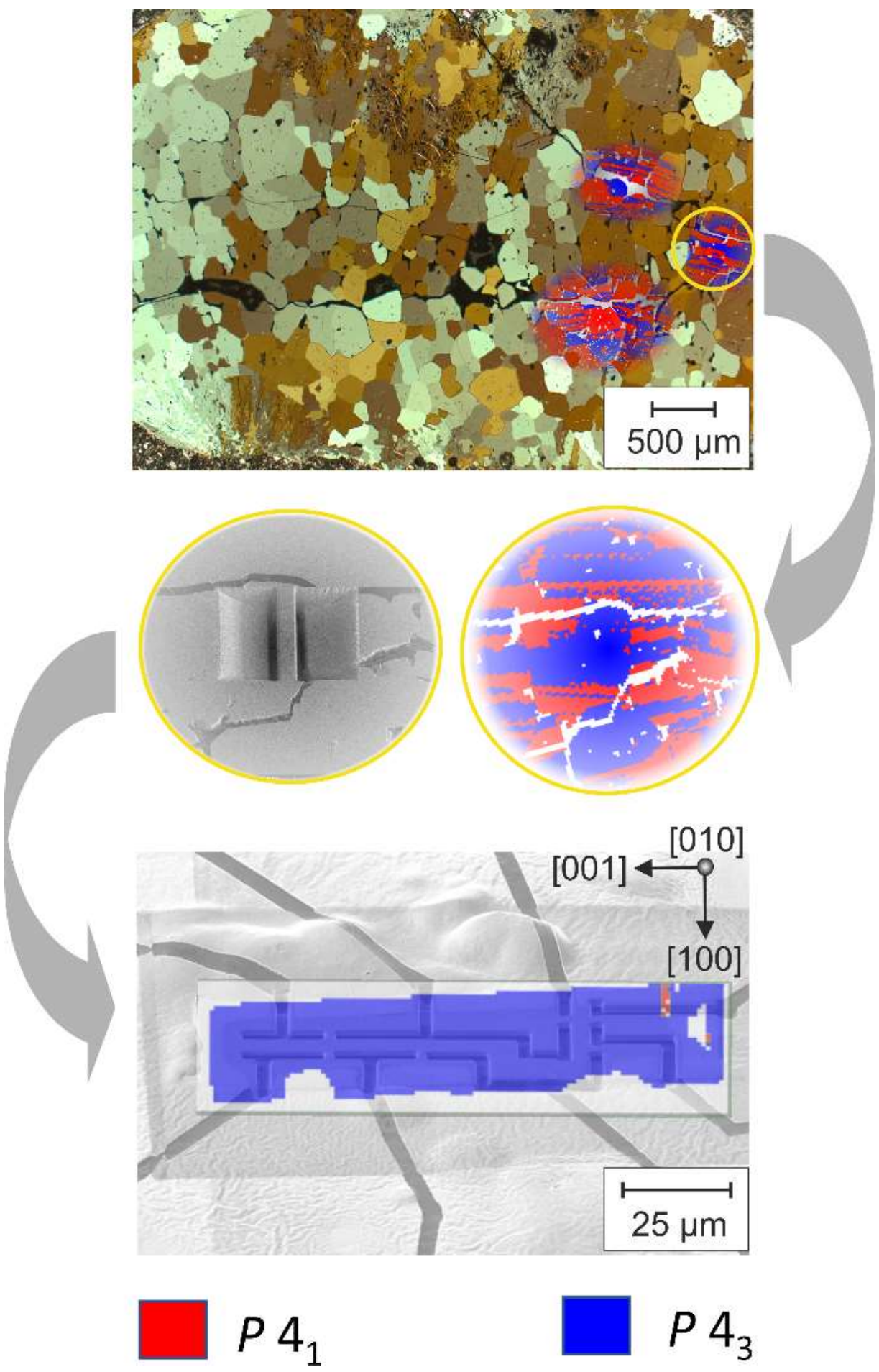

**Fig. 1.** $LaRhC_2$: a schematic representation for the microdevice preparation. *Top*: optical micrograph (in polarized light) of the polished surface of a polycrystalline material. The grains show strong orientation contrast (cavities and cracks are black). Red and blue colors indicate respective handedness – right-handed (red; space group $P4_1$) and left-handed (blue, space group $P4_3$) monochiral domains, according to EBSD. *Middle*: FIB cut of a lamella from a monochiral surface area, the lamella is oriented parallel to the fourfold screw axis. *Bottom:* a ready-to-measure transport microdevice with well-defined voltage probes.

**Crystal structure determination and thermal expansion behaviour**

Two FIB-cut cubes of $LaRhC_2$ with different handedness (Fig. 2) were investigated by scXRD experiments. Details of the data collection and results of the structural refinements are listed in Tables 1, S4-S7. Assuming the space groups assigned by EBSD, we used the direct methods to localize the crystallographic positions for La and Rh. The coordinates of carbon atoms were obtained by the difference Fourier maps. Further, weighting scheme, extinction coefficients, atomic coordinates (coordinate $z$ for La in each case was fixed at $z$ = 0, for reference), anisotropic ($U_{ij}$) as well as isotropic ($U_{\mathrm{iso}}$) displacement parameters for La, Rh, and for C1, C2-atoms were refined (Tables 1 and S4-S6). The crystal structure refinement confirmed the structure model of $LaRhC_2$ and agreed with the EBSD-based assignment of the handedness for both crystals. One peculiarity of the crystal structure is an occurrence of the $C_2$-dumbbells ($d_{\mathrm{C-C}} \approx 1.36(1)$ Å, Table S7) with a distance close to a C=C bond ($d_{\mathrm{C-C}} = 1.34$ Å), like in olefins (atomic radii $r_{\mathrm{C}} = 0.77$ Å)[41]. The $C_2$-dumbbell is embedded in the tricapped trigonal prism [$La_6Rh_3$]. These units form right- and left-handed spirals along the $4_1$ and $4_3$ screw axes, respectively (Fig. 2c), by sharing their base and trigonal faces.

ScXRD data collection at 200 K and 100 K showed the anisotropic temperature dependence of the crystal structure which led to an increase in the $c/a$ ratio with decreasing temperature (Fig. 3). The detailed analysis of the lattice parameters, thermal displacement parameters and interatomic distances revealed the same behavior for both enantiomorphs (Tables S4-S7). The corresponding scXRD crystallographic data can be obtained from Cambridge Crystallographic Data Centre (https://www.ccdc.cam.ac.uk/structures, CCDC 2413251–2413256).

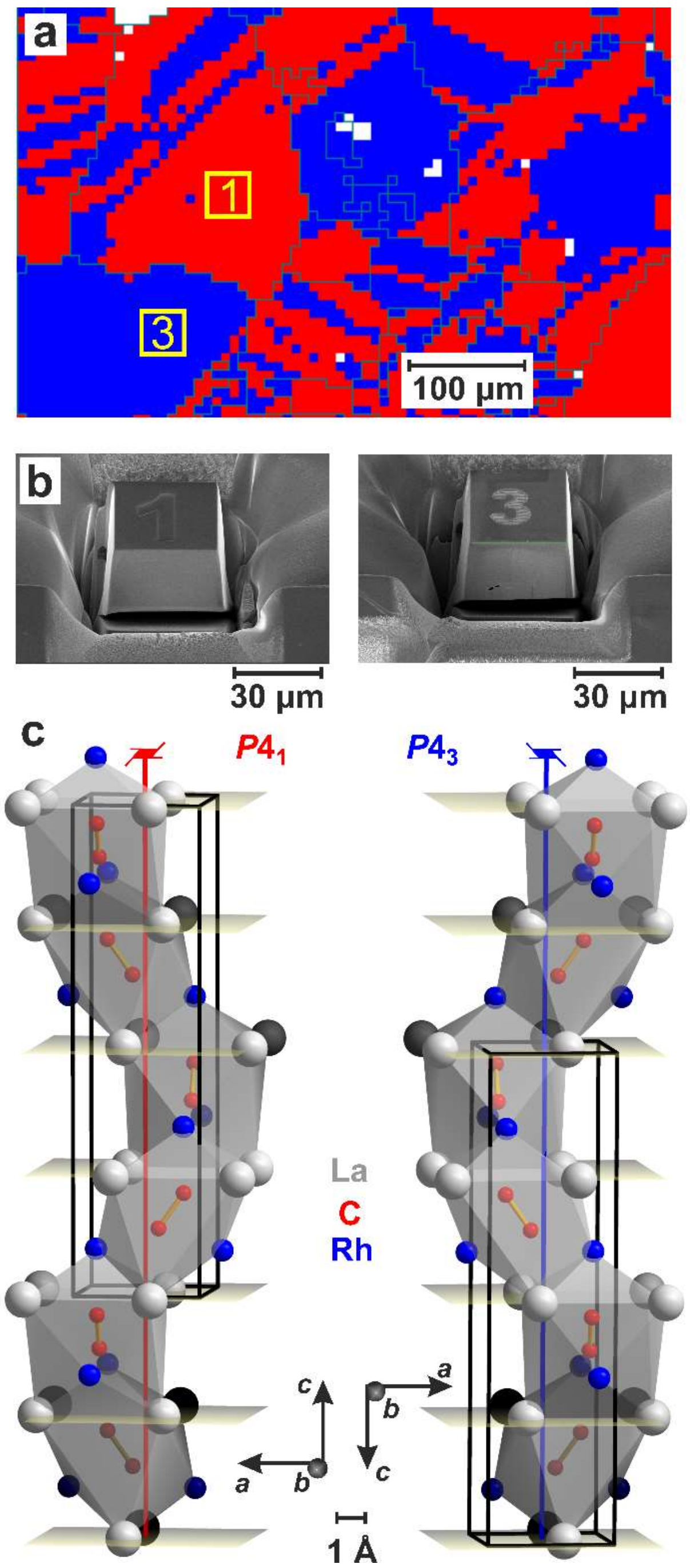

**Fig. 2.** Chirality analysis of $LaRhC_2$. **a)** Enantiomorph-distribution map of the selected region of a polycrystalline $LaRhC_2$ sample (red colour corresponds to $P4_1$-$LaRhC_2$, blue – $P4_3$-$LaRhC_2$) obtained from EBSD. Extraction areas of the single crystals are marked by numbers. **b)** FIB-cut single-crystalline cuboids of $P4_1$ (1) and $P4_3$ (3) enantiomorphs. **c)** Arrangement of the coordination polyhedra of the $C_2$-dumbbells along the screw axis of the enantiomorphs.

**Table 1 | Details on the $LaRhC_2$ single crystal XRD data collection and analysis**

| | | |
|---|---|---|
| Temperature (K) | 293(2) | |
| Space group | $P4_1$ | $P4_3$ |
| Specimen shape | cuboid | |
| Crystal size (min-mid-max, µm) | 35×37×42 | 32×38×42 |
| Unit cell parameters $a$, $c$ (Å) | 3.9703(1), 15.3323(3) | |
| $hkl$ range | $-6 \leq (h, k) \leq 6$, $-25 \leq l \leq 25$ | |
| No. of measured reflections | 24510 | 24695 |
| No. of independent reflections | 1150 | 1150 |
| No. of reflections with $I > 2\sigma(I)$ | 1029 | 999 |
| No. of refined parameters | 27 | 27 |
| $R_{int}$ (%) | 5.1 | 5.3 |
| $R_F$ (%) | 1.6 | 1.7 |
| $wR$ (%) | 3.5 | 3.8 |
| *Goodness-of-fit* | 1.10 | 1.09 |
| No. of Bijvoet pairs | 465 | 460 |
| Flack parameter | −0.02(2) | −0.02(2) |
| Residual electron density (max/min, $e^-$ Å$^{-3}$) | +0.82/−1.00 | +0.81/−1.28 |

A detailed analysis of the anisotropic temperature dependence of the UCPs of $LaRhC_2$ in the temperature range 100 K < $T$ < 293 K allowed refinements based on high-resolution synchrotron powder XRD data (Table S8). The decrease of the lattice parameters and the volume of the tetragonal unit cell with decreasing temperature is accompanied with the increase in the $c/a$ ratio (Fig. 3). There were also differences in the thermal expansion at ambient temperatures $\alpha_a = 8.8\times10^{-6}$ K$^{-1}$ and $\alpha_c = 5.2\times10^{-6}$ K$^{-1}$ with a tendency to become more isotropic at lower temperatures ($T$ = 100 K: $\alpha_a = 5.2\times10^{-6}$ K$^{-1}$, $\alpha_c = 2.8\times10^{-6}$ K$^{-1}$; insets of Fig. 3). The thermal expansion coefficients of $LaRhC_2$ at ambient temperature are comparable to those of many hard metals as Ce ($6.3\times10^{-6}$ K$^{-1}$), Pr ($6.7\times10^{-6}$ K$^{-1}$), Hf–Pt (4.5-8.8 $\times10^{-6}$ K$^{-1}$).[42]

This anisotropy and its temperature dependence can be attributed to both enantiomorphs because left- and right-handed contributions resulted the same diffraction pattern in the investigated temperature range.

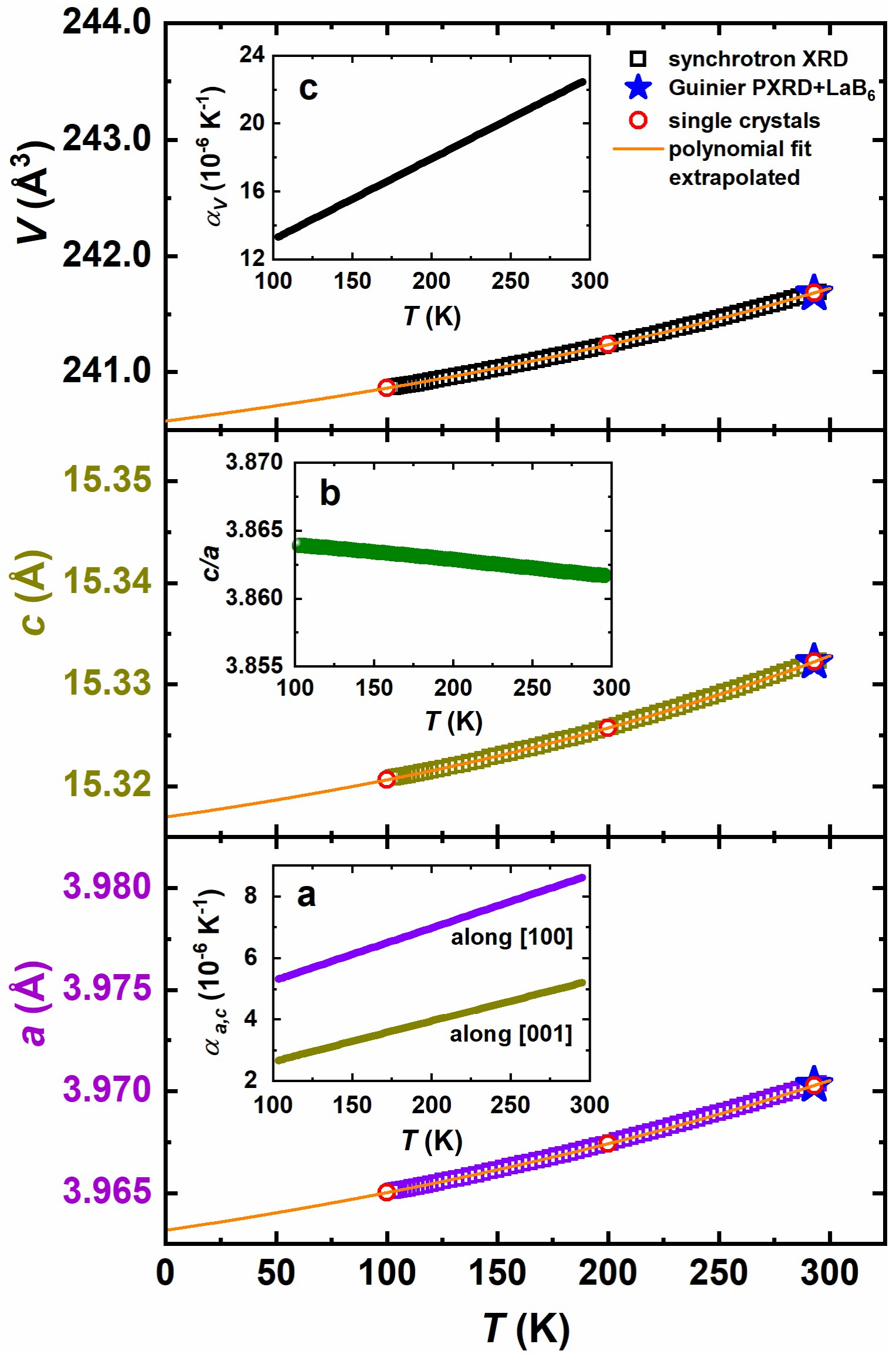

**Fig. 3.** Temperature dependence of the tetragonal unit cell of $LaRhC_2$: UCPs *a*, *c* and volume *V*, orange solid lines are fits to a quadratic polynomial. **Insets: a)** linear thermal expansions along [100] and [001], **b)** *c*/*a* ratio, **c)** volume thermal expansion coefficient.

**Electronic structure calculations**

The electronic band structure calculations for $LaRhC_2$ were performed using the extrapolated to $T = 0$ K UCPs (Fig. 4, Table S8), with further relaxation of atomic coordinates refined from scXRD (Tables S5, S6). It shows an indirect band gap with $E_g^{DFT} \approx 75$ meV (inset of Fig. 4). The electronic density of states for $LaRhC_2$ indicates that the valence band (below the Fermi level $E_F$) consists of three separate energy regions centered at $E \approx -8$ eV, $-5.5$ eV and $-2$ eV, respectively (Fig. 4). The lowest energy region is composed of Rh-4*d*, C-2*p* and C-2*s* states (not shown in Fig. 4). The energy region with the center at $E \approx -2$ eV is dominated by C-2*p* with admixtures of Rh-4*d* and La-5*d* states. Finally, the broad band extending from $E_F$ down to $E \approx -4$ eV is mainly due to Rh-4*d*, C-2*p* and La-5*d* states. The performed calculation indicates that La-5*d* and La-6*s* (not shown in Fig. 4) states to be almost unoccupied (i.e. are situated above $E_F$), which hints at a charge transfer towards the negatively charged $[RhC_2]_n^{m-}$ framework. Both the enantiomorphs have equivalent band structure (Fig. S4) and thus are characterized by similar fat-band diagrams (Figs. S4, S5).

The 1st Brillouin zone of the $LaRhC_2$ together with the marked high-symmetry points is drawn in Fig 4b. Interestingly, the conduction valence band maximum (VBM) is located at X point, and the conduction band minimum (CBM) is occurring on the path X–M, closer to M(½, ½, 0) [inset of Fig. 4 and Fig. 4b]. The superimposed Fermi surfaces of these bands in the close vicinity of the Fermi level (pockets) is shown in Fig. 4b. Noteworthy, the CBM surface is of quite regular form, almost spherical, while the ellipsoidal surface of the VBM is strongly elongated along $z^*$, also non-isotropic in the plane $\perp z^*$, being more expanded in the X–Γ direction. Such strong anisotropy in the band structure is indicative of a significant anisotropy in the electrical conductivity parallel and perpendicular to the fourfold screw axis, as well as variation for different directions perpendicular to the fourfold screw axis.

In general, our calculations are comparable with those recently published[14] with an indirect band gap of $E_{\mathrm{g}}^{\mathrm{DFT}} \approx 84$ meV. Contrary, the electronic band structure calculations obtained by applying the PBE exchange correlation functional (*WIEN2k* code)[17] indicate a semi-metallic scenario.

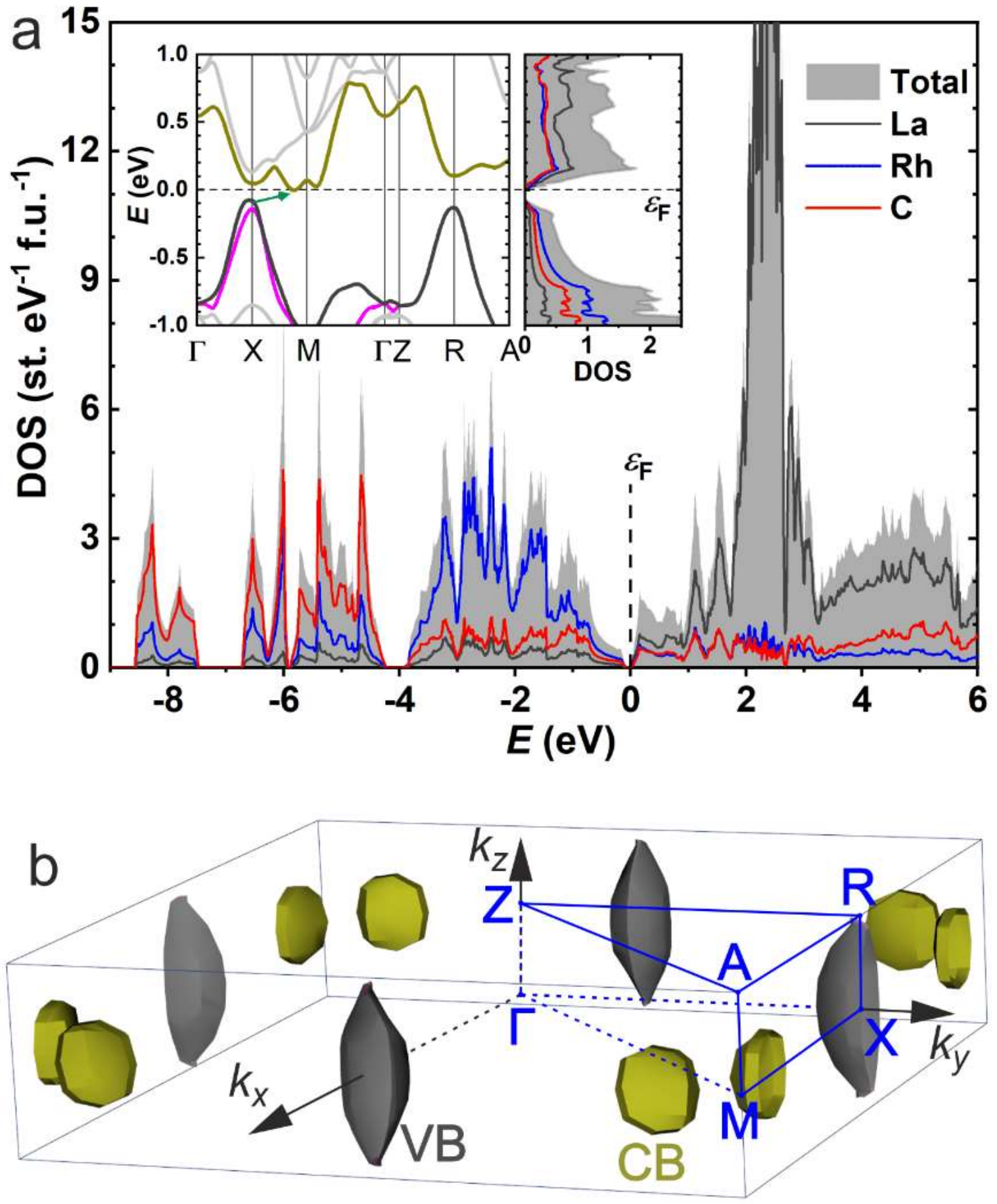

**Fig. 4.** Electronic structure of $LaRhC_2$. **a)** Electronic DOS together with element-resolved DOS contributions. **Inset:** electronic band structure and DOS near the Fermi level. An indirect energy gap of ~75 meV is marked by a green arrow. **b)** 1st Brillouin zone of $LaRhC_2$ with indicated high-symmetry points filled with the superimposed Fermi surfaces of the respective conduction (CB) and valence (VB) bands near the Fermi level.

**Electrical transport measurements**

The temperature dependence of electrical resistivity in the 1.8 K < $T$ < 300 K temperature range shows semiconducting behavior perpendicular [$\rho_{xx}^{\perp}(T)$] as well as parallel [$\rho_{xx}^{\parallel}(T)$] to the 4-fold screw axis (Fig. 5) with significant orientation dependence. The measurements indicate that $\rho_{xx}^{\parallel}(T) > \rho_{xx}^{\perp}(T)$, which correlates well with the structural characteristics for $c > a$ (Table 1), as for example is the case for crystals with tetragonal or hexagonal symmetry.[43,44,45] Despite visually different curvature of $\rho_{xx}^{\parallel}(T)$ and $\rho_{xx}^{\perp}(T)$, the relative increase at 1.8 K compared to RT values [alternative to the $RRR$ (residual resistivity ratio) for metallic samples] for most curves is nearly the same ≈3.2–3.9. The only exception is $\rho_{xx}^{\perp 4_3}(T)$. Its unusual behavior is correlated with the alignment of [110] parallel to the electrical current, which is different to the measurement along [100] for the specimens with the $P4_1$ symmetry.

By evaluating the exponential behavior (Arrhenius approximation) of $\rho_{xx}(T)$ in the 100 K < $T$ < 300 K temperature range, slightly different values of the energy gap $E_{\mathrm{g}}^{\rho}(\parallel) \approx 20$ meV and $E_{\mathrm{g}}^{\rho}(\perp) \approx 32$ meV are obtained (Table S9). This can be explained by the non-isotropic thermal expansion ($\alpha_a > \alpha_c$), leading to the deformation of the band structure with changing temperature. The estimated band gaps parallel to the fourfold screw axes are qualitatively the same for both enantiomorphs, while the small difference in the case of $E_{\mathrm{g}}^{\rho}(\perp)$ can be explained by a difference in the [100] and [110] crystallographic orientation for $P4_1$ and $P4_3$ microdevices, respectively (Table S9).

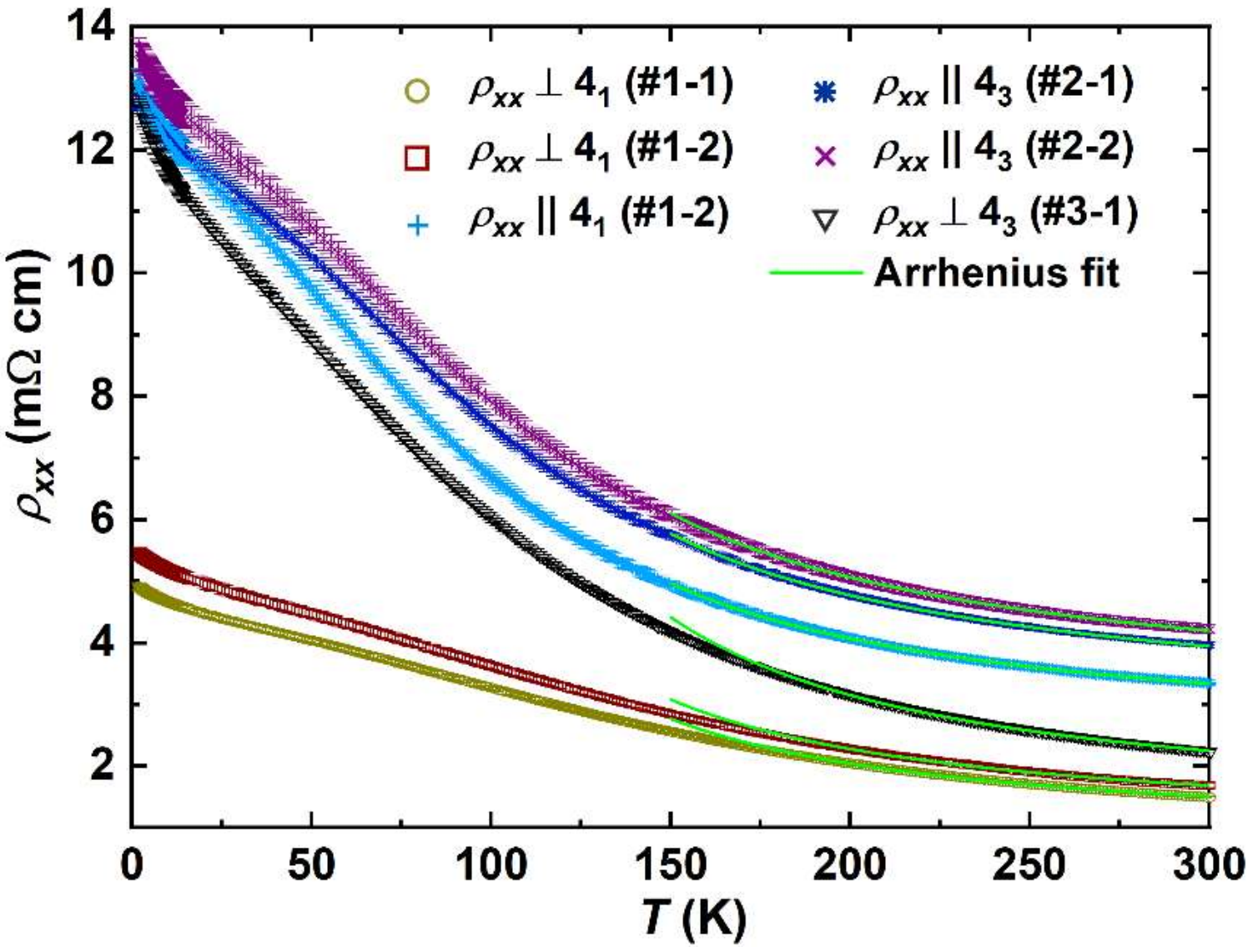

**Fig. 5.** Temperature-dependent longitudinal electrical resistivities $\rho_{xx}$, probed parallel (||) and perpendicular (⊥) to the 4-fold screw axes together with the Arrhenius approximation fits (lines). The microdevices are manufactured from three different grains of the polycrystalline sample [$P4_1$-$LaRhC_2$ (#1) and $P4_3$-$LaRhC_2$ (#2 and #3)].

To determine the Hall coefficients $R_H$, the transverse ($\rho_{xy}(H)$) electrical resistivities were measured in magnetic field ($H$). The respective Hall isotherms are shown in Fig. S6. Up to 10 T and down to ~15 K, they exhibited linear dependencies with slopes corresponding to $R_H$; the latter are plotted in Fig. 6a as $R_H(T)$. The Hall coefficients are negative in all studied temperature ranges, indicating $n$-type conductivity. The charge carrier concentrations (Fig. 6b) were calculated from: $n = \left|\frac{1}{eR_H}\right|$ ($e$ is the elementary charge). In the 15–100 K temperature range, $n$ for all samples is nearly temperature-independent, with and of $10^{19}$ $cm^{-3}$ order of magnitude. High charge carrier concentration of this order is usually observed in the heavily-doped semiconductors.[46] For $T > 100$ K, $n(T)$ increases drastically with temperature, following $n(T) \propto T^{3/2}$ dependence (Fig. 6b). Such behavior would indicate a transition into the intrinsic regime.

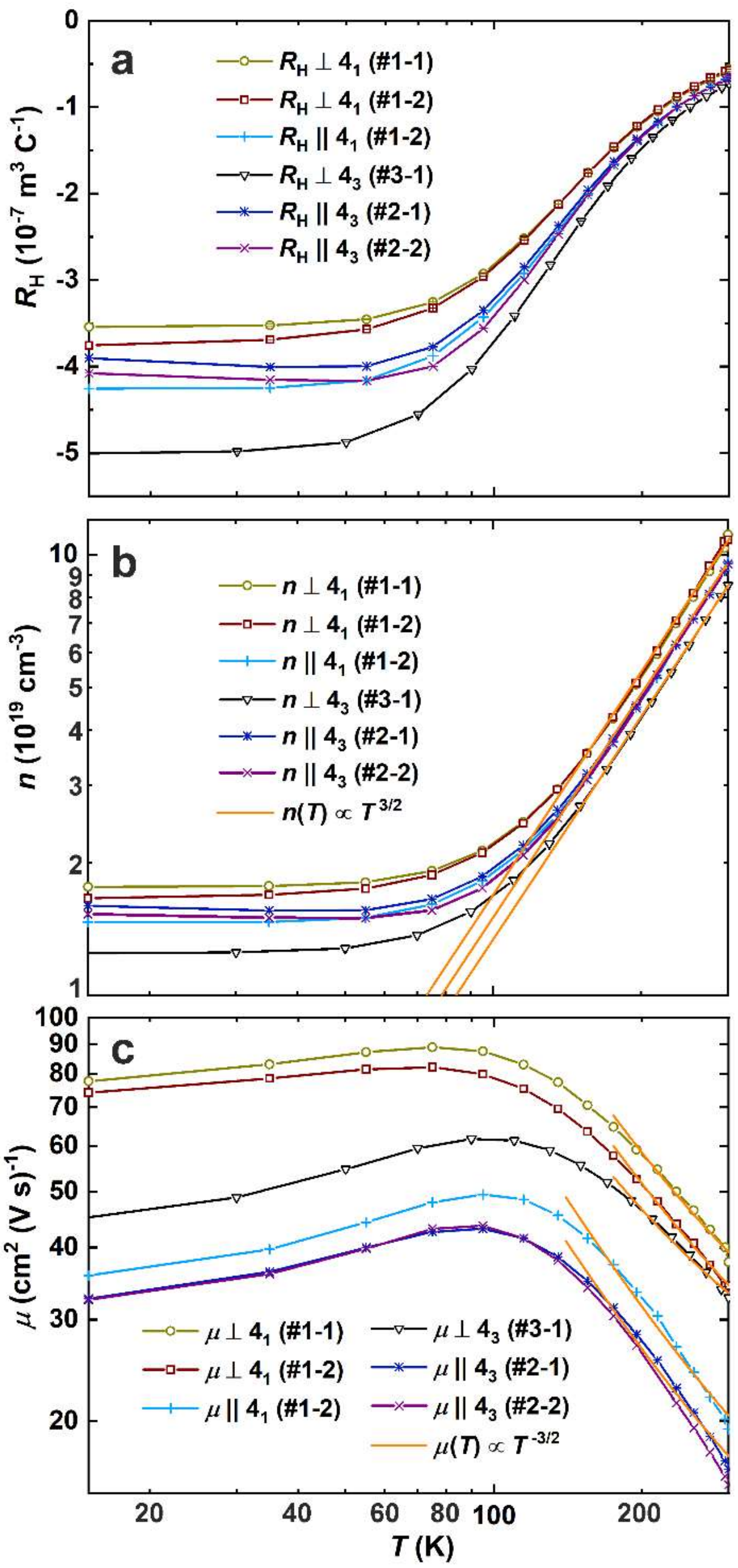

**Fig. 6.** $LaRhC_2$-charge carriers' characteristics. **a)** Enantiomorph-resolved Hall coefficients. **b)** Charge carrier concentrations, with orange lines indicating the $n(T) \propto T^{3/2}$ dependencies. **c)** Charge carriers mobilities, together with characteristic power *T*-dependencies (orange lines).

From the longitudinal resistivities and Hall coefficients, we calculate the charge carrier mobilities as $\mu = \frac{1}{ne\rho_{xx}(T)} = \left|\frac{R_{\mathrm{H}}}{\rho_{xx}(T)}\right|$ (Fig. 6c). In general, the absolute values of mobilities are moderately low, and increasing insignificantly, due to high charge carrier concentration, with increasing temperature. At $T \approx 100$ K, $\mu(T)$ starts to decrease as $\propto T^{-3/2}$, implying the dominate scattering by acoustic phonons.

The temperature dependence of the carrier concentration is virtually identical for all measured specimens, with a slightly higher concentration, measured parallel to [100]. The concentration is the smallest for the specimen with [110] orientation and $P4_3$ symmetry.

**Anisotropic magnetoresistance**

The angle-dependent measurements of the longitudinal magnetoresistance with a 10 T magnetic field oriented parallel or antiparallel to the fourfold screw axis have been performed in the 10 K < $T$ < 330 K temperature range (0° ≤ $\varphi$ ≤ 360°, where $\varphi$ – is an angle between the direction of electric current ($\vec{j}$) and direction of magnetic field strength ($\vec{H}$)). Selected curves of the absolute magnetoresistance calculated as $\Delta\rho_{xx}^{10\mathrm{T}} = \rho^{10\mathrm{T}} - \rho^{0\mathrm{T}}$ are presented in Fig. S3 (for ease of viewing, error bars are not shown). In general, no remarkable handedness dependence is observed, with the data being highly symmetric despite weak signal (it is worth noting, that the calculated magnetoresistance is reaching maximum ≈1% of the respective conventional electrical resistivity: $MR = \frac{\rho^{10} - \rho^{0T}}{\rho^{0T}} \times 100\%$). Interestingly, by plotting $\Delta\rho_{xx}^{0^\circ} - \Delta\rho_{xx}^{180^\circ}$ against the absolute magnetic field, different slopes can be deduced for the opposite handed $LaRhC_2$ enantiomorphs (see Fig. S3 for the dataset at $T_5$ = 10 K). This indicates the occurrence of a weak electromagnetic chiral anisotropy (eMChA), which is also observed in Bi-wires with opposite helix dislocations[47], as well as in the case of chiral carbon nanotubes[48]. These findings agree with the ideas of Rikken & Raupach who suggested the handedness impact on the electromagnetic transport under aligned external magnetic field[49].

At high temperatures down $T \approx 200$ K, $MR$ is practically absent when $\vec{j} \parallel \vec{H}$, and the largest values are observed for $\vec{j} \perp \vec{H}$ (Fig. S3). Thus, both $LaRhC_2$ enantiomorphs reveal weak anisotropic magnetoresistance, normally observed in low-dimensional or highly anisotropic nonmagnetic crystalline materials as e.g. β-$Ag_2Te$[50] or $PdSe_2$[51].

## Summary and conclusions

Although chirality is commonly linked to topological materials like Weyl semimetals and chiral superconductors, its impact on charge transport also manifests in topologically trivial systems. In this work we examine $LaRhC_2$, which forms in the chiral crystal structures described by space groups $P4_1$ and $P4_3$. The chemical and physical properties of single-crystalline enantiomorphs were studied separately in order to access the role of chirality. The high-quality micro-scale specimens and devices were prepared by means of EBSD and FIB techniques, with in-depth studies of their structural and electrical transport properties. The $P4_1$-$LaRhC_2$ and $P4_3$-$LaRhC_2$ are shown to be the narrow-band semiconductors with same electronic band structure, which agree with the theoretical analysis. Both enantiomorphs display similar absolute values and temperature dependencies of the electrical conductivity along the fourfold screw axes. A weak anisotropic magnetoresistance is revealed. $\rho_{xx}(T)$ dependencies perpendicular to screw axes indicated well pronounced anisotropic behaviors with respect to the crystallographic direction of the oriented crystal. Overall, this study serves as a foundational step toward understanding chirality-dependent properties in low-symmetry quantum materials. The integrated use of EBSD and FIB microfabrication techniques establishes a methodological benchmark for investigating the role of chirality in charge transport.

## Acknowledgments

This work was performed within the DFG project 467257848. The DynaCool-12 device was inquired within the DFG project 422219907. E.S. is grateful for the support of the Christiane Nusslein-Volhard-Stiftung and the Boehringer Ingelheim Plus 3 Program. We are indebted to H. Borrmann for the help with the PXRD experiments, to Silvia Kostmann for her assistance during metallographic preparations and analyses. We thank F. Meier, and A. Schwarzer for the help with scXRD experiments. Special thanks go to A. Winkelmann for advice and discussion on the EBSD pattern analysis.

# Supplementary materials

*for*

# Anisotropy of the chiral, semiconducting phase $LaRhC_2$: a handedness resolved study

Volodymyr Levytskyi[1,*], Ulrich Burkhardt[2,*], Markus König[2], Christoph Hennig[3,4], Eteri Svanidze[2], Yuri Grin[2], and Roman Gumeniuk[1,*]

[1] Institut für Experimentelle Physik, TU Bergakademie Freiberg, Leipziger Str. 23, 09599 Freiberg, Germany

[2] Max-Planck-Institut für Chemische Physik fester Stoffe, Nöthnitzer Str. 40, 01187 Dresden, Germany

[3] Helmholtz-Zentrum Dresden-Rossendorf, Institute of Resource Ecology, Bautzner Landstr. 400, 01314 Dresden, Germany

[4] Rossendorf Beamline (BM20-CRG), European Synchrotron Radiation Facility, 71 Avenue des Martyrs, 38043 Grenoble, France

*Corresponding authors' E-mails: volodymyr.levytskyi@physik.tu-freiberg.de
ulrich.burkhardt@cpfs.mpg.de
roman.gumeniuk@physik.tu-freiberg.de

**Synthesis optimization**

To obtain better quality samples, the following procedure was experimentally deduced. First, the phase analysis performed on the PXRD pattern of the stoichiometric sample obtained directly after arc-melting revealed the majority phase $LaRhC_2$ as well as the minority phases $LaRh_2$, $LaC_2$, free carbon and additional unindexed diffraction lines (Fig. S1a). Further annealing of the alloy at 1070 K increase the amount of $LaRhC_2$ (Fig. S1a and Table S1), which prompted us to perform a heat treatment at 1520 K. As one can see from Fig. S1a,b, the obtained after such a procedure PXRD pattern of the sample contained only minor intensities belonging to $LaRh_2$ (<1 wt. %, Table S1) as well as six unindexed weak peaks. Importantly, the full width at the half maxima (FWHM) of these reflections differ strongly from those of the targeted $LaRhC_2$ phase (inset to Fig. S1b), which let us assume that they are unlikely to indicate a superstructure. The stoichiometric sample contained the least amount of impurities as follows from PXRD.

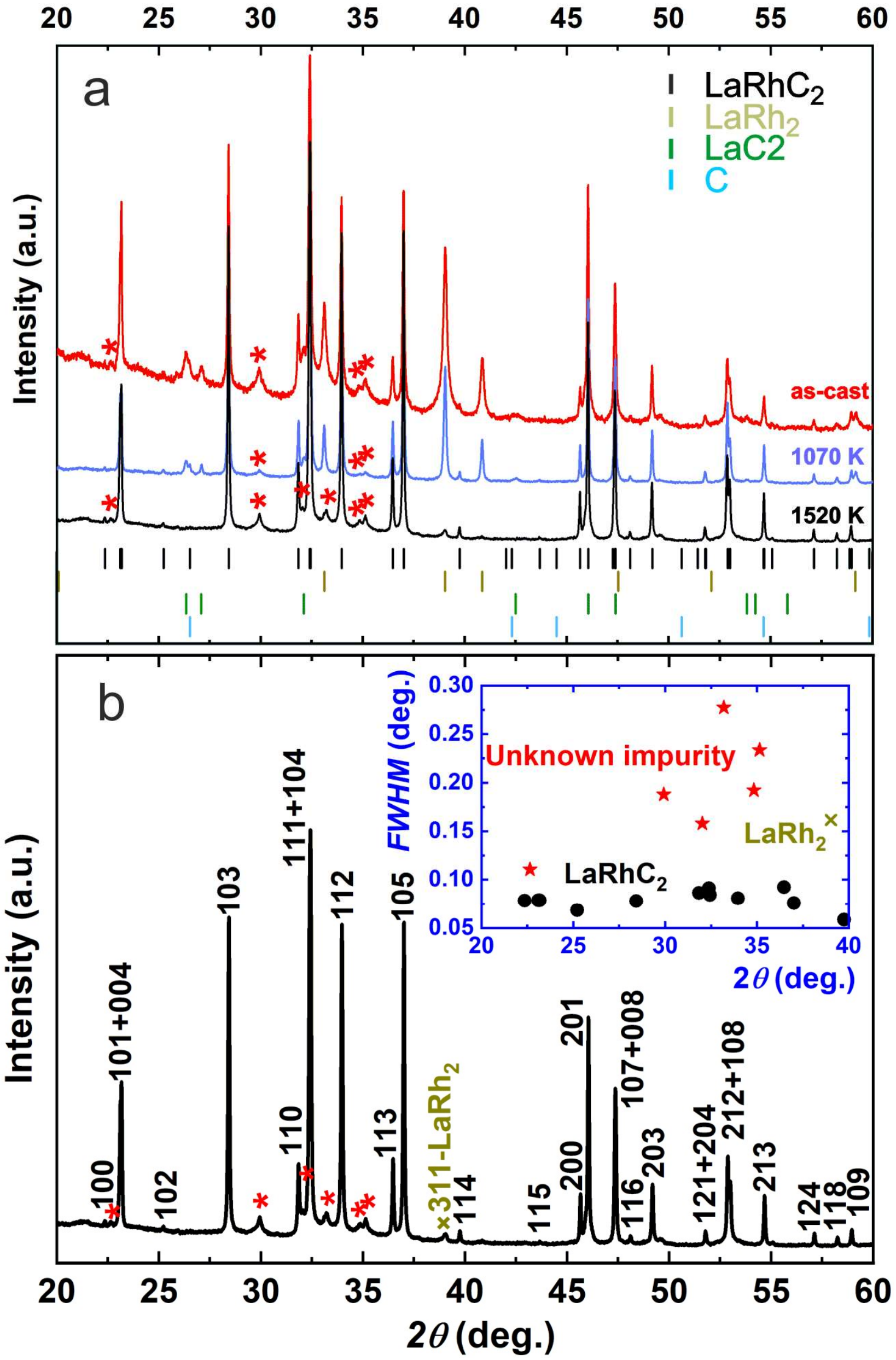

**Figure S1. a)** Indexing of PXRD patterns of the as-cast and annealed at different temperatures $LaRhC_2$ sample. Theoretical peak positions of the identified phases are marked by vertical bars. **b)** Indexed PXRD pattern of the annealed at 1520 K polycrystalline $LaRhC_2$ sample. Peaks belonging to the unidentified impurity are marked by red asterisks. **Inset:** Full width at half maxima of the diffraction peaks of main and impurity phases.

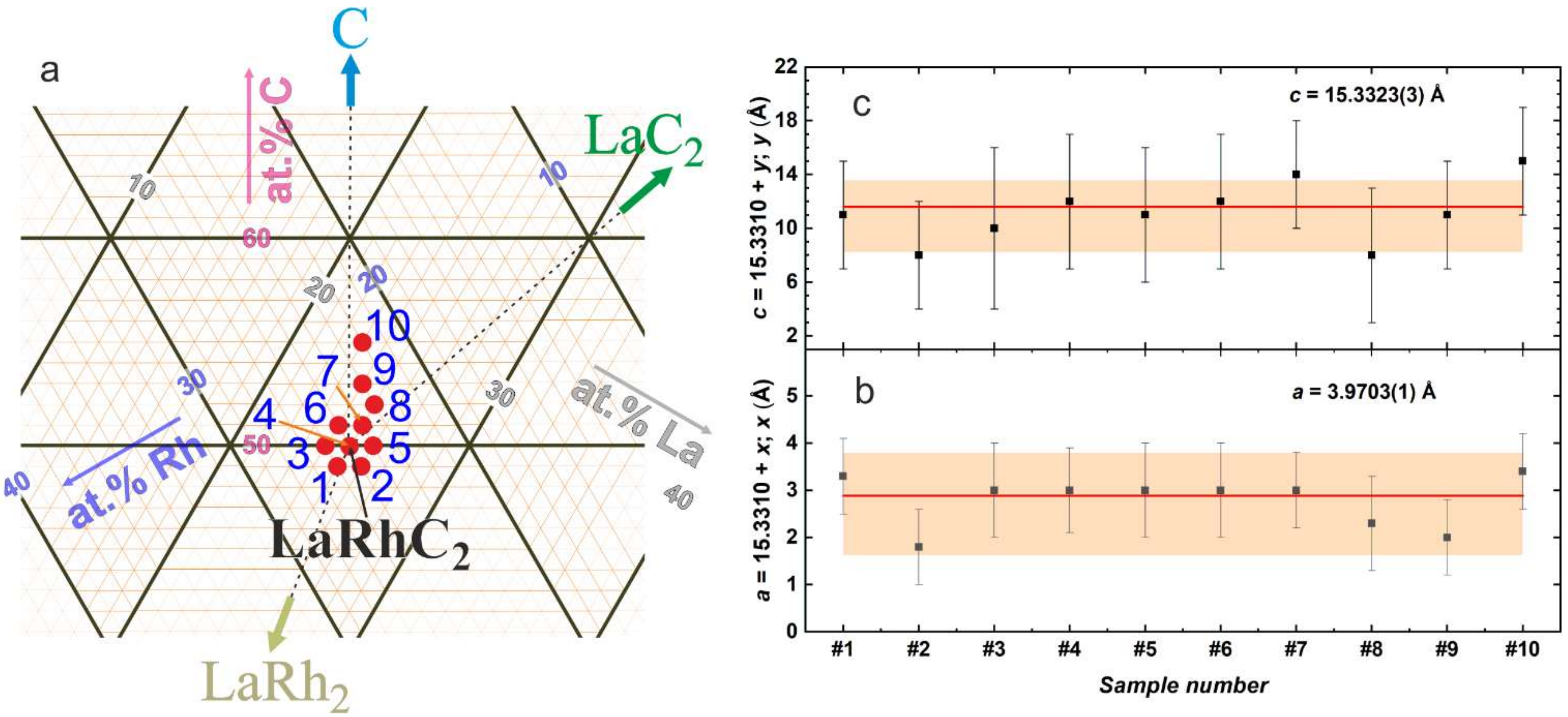

**Figure S2. a)** Selected area of the Gibbs-Rosebom triangle with the depicted starting chemical compositions of the synthesized La-Rh-C samples. **b, c)** UCPs of the $LaRhC_2$ phase determined by PXRD using $LaB_6$ as internal reference in the annealed samples.

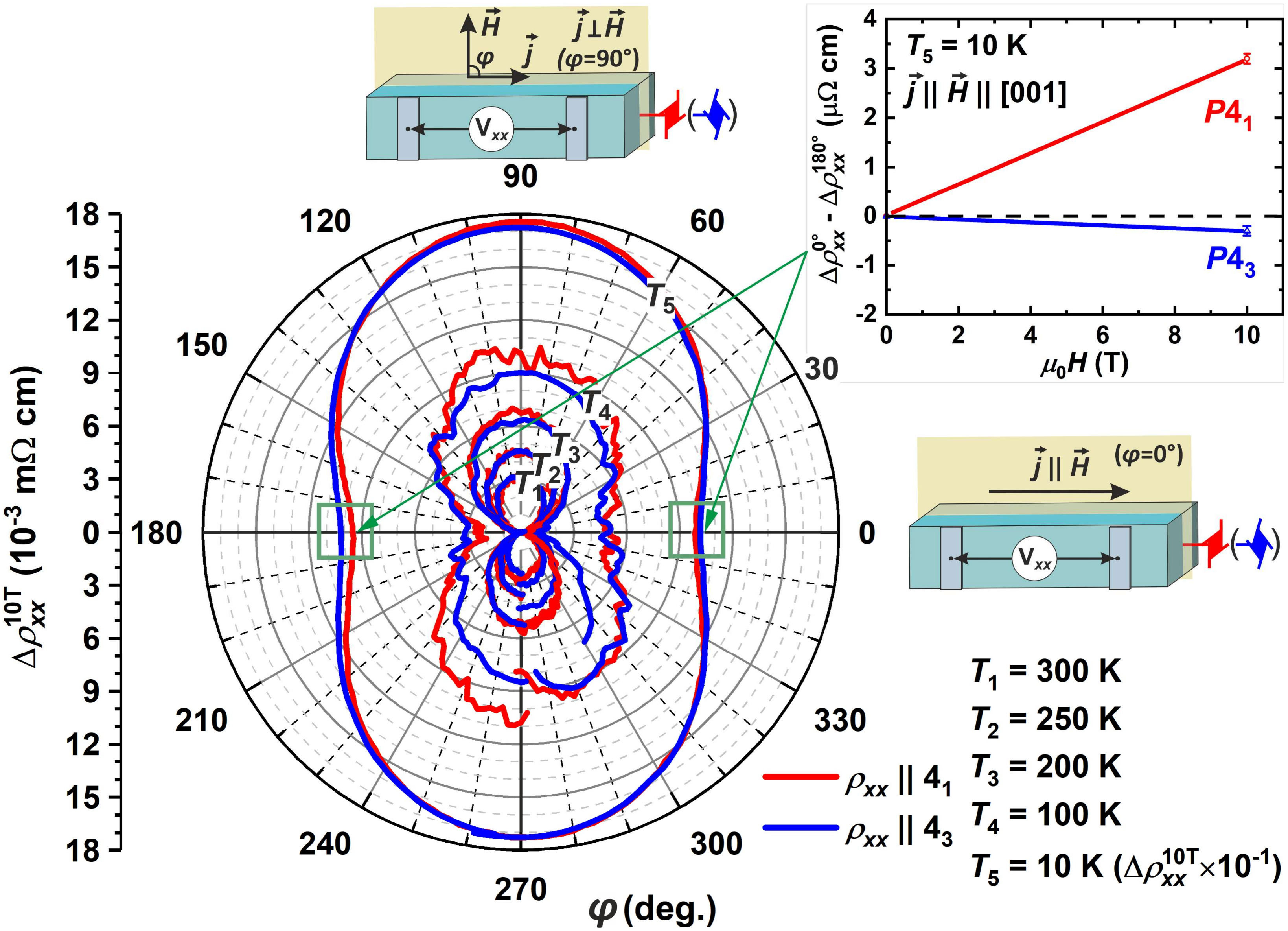

**Figure S3.** Angle-dependent magnetoresistance of $P4_1$- and $P4_3$-$LaRhC_2$ at oriented magnetic field of 10 T, obtained as $\Delta\rho_{xx}^{10T} = \rho_{xx}^{10T} - \rho_{xx}^{0T}$ at various temperatures.

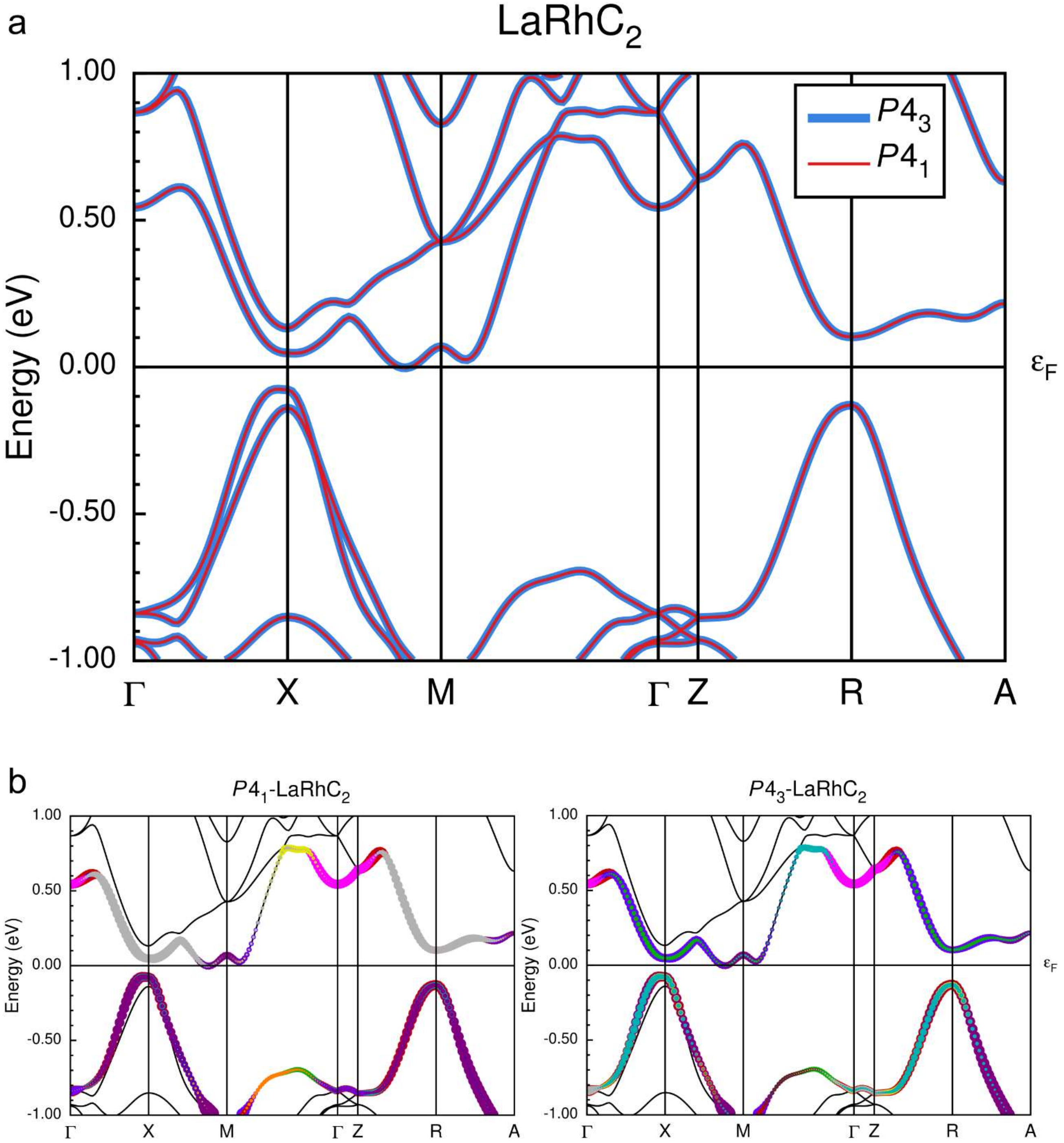

**Figure S4. a)** Electronic band structure of the right-handed (SG $P4_1$) and left-handed (SG $P4_3$) $LaRhC_2$ near the Fermi level ($\varepsilon_F$). **b)** Fat-band diagrams of the upper valence and lower conduction bands (VB: Rh-4*d*, C-2*p*; CB: La-5*d*, Rh-4*d*, C-2*p*; weight coefficient 10) of the right-handed and left-handed $LaRhC_2$, accordingly.

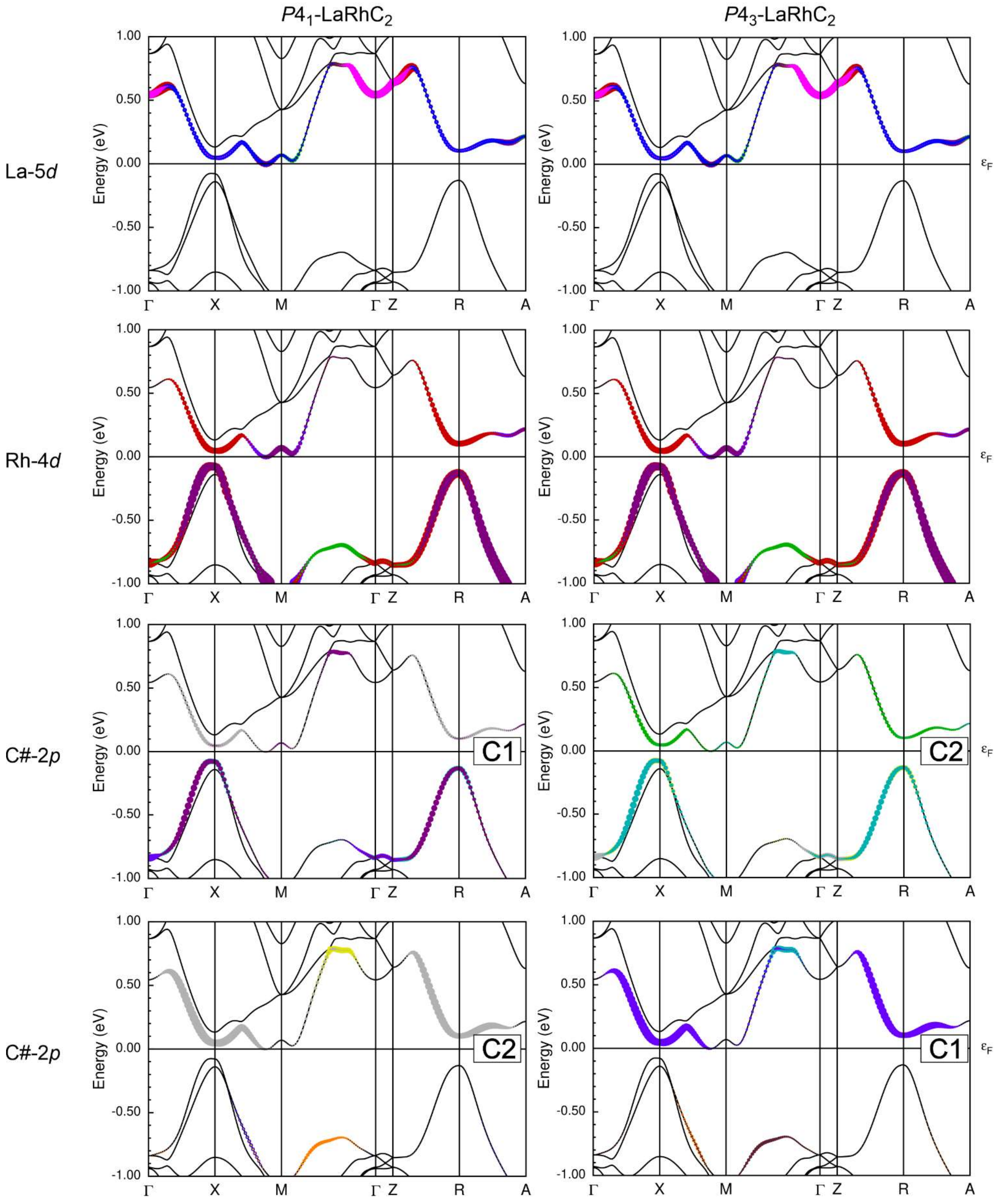

**Figure S5.** Element-resolved fat-band diagrams of the upper valence and lower conduction bands (weight coefficient 10) of the right-handed (SG $P4_1$) and left-handed (SG $P4_3$) $LaRhC_2$, accordingly.

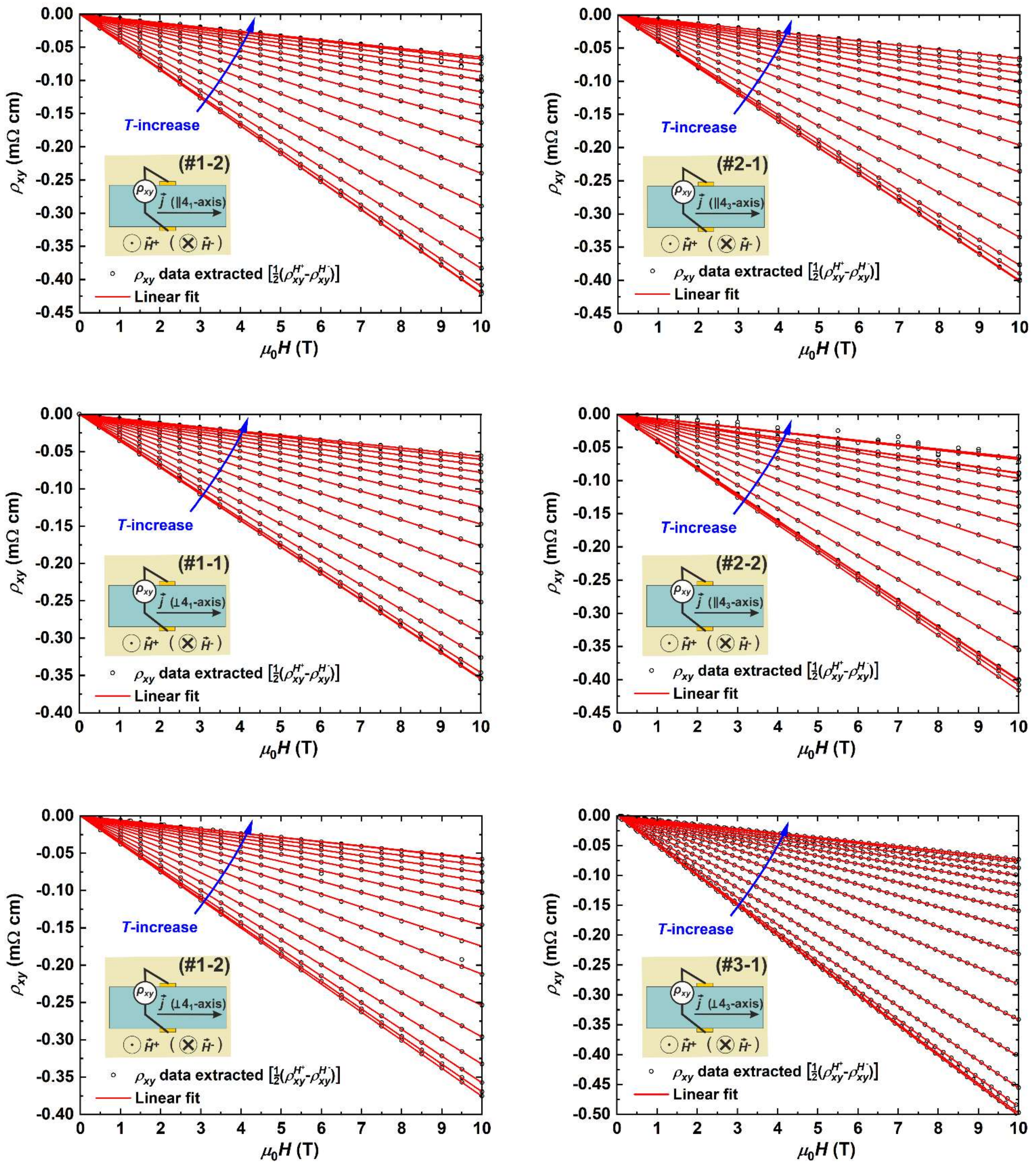

**Figure S6.** Hall isotherms ($\rho_{xy}$ vs. $H$) measured for the selected microdevices (for geometrical parameters, see Tables S2, S3). The schematic arrangement of the electrical contacts is shown in insets.

**Table S1.** Results of the phase analysis by profile fitting of the as-cast and annealed samples of the stoichiometric initial composition $LaRhC_2$.

| **Sample** | **Profile factor, $R_p$ (%)** | **wt. % $LaRhC_2$** | **wt. % $LaRh_2$** | **wt. % $LaC_2$** | **wt. % C** |
|---|---|---|---|---|---|
| As-cast | 3.7 | 72(3) | 20(2) | 4(2) | 4(2) |
| Annealed at 1070 K | 5.3 | 78(2) | 16(2) | 2(1) | 4(1) |
| Annealed at 1520 K | 3.8 | 99(4) | 1(1) | – | – |

**Table S2.** Measured $P4_1$-$LaRhC_2$ microdevices with the final handedness mapping (from EBSD)*, geometrical parameters of the bars [voltage separation probes length ($l$) and cross-sectional areas ($S$)], and practical orientation of the fourfold screw relative to the current flow and the lamella surface.

| **#1-1: $(\rho_{xx}, \rho_{xy}) \perp 4_1$** | **#1-2: $(\rho_{xx}, \rho_{xy}) \perp 4_1$** | **#1-2: $(\rho_{xx}, \rho_{xy}) \parallel 4_1$** |
|---|---|---|
| 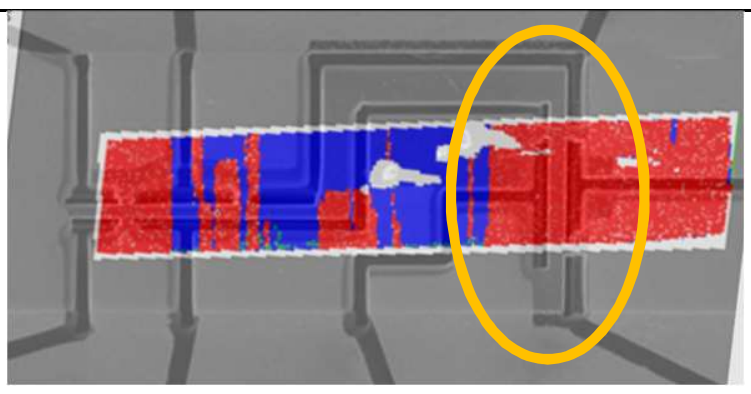 | 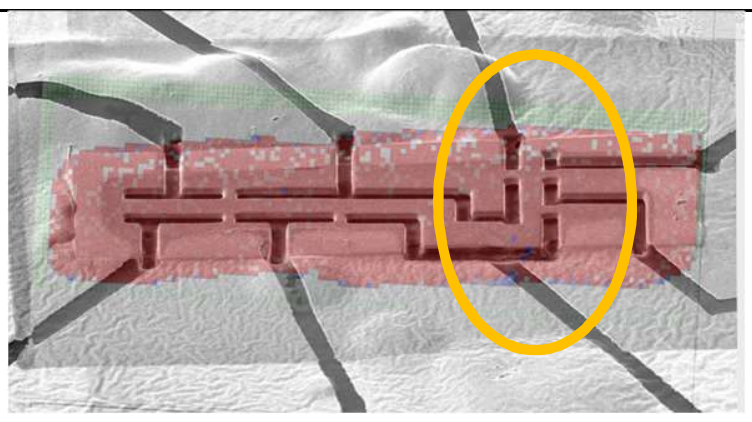 | 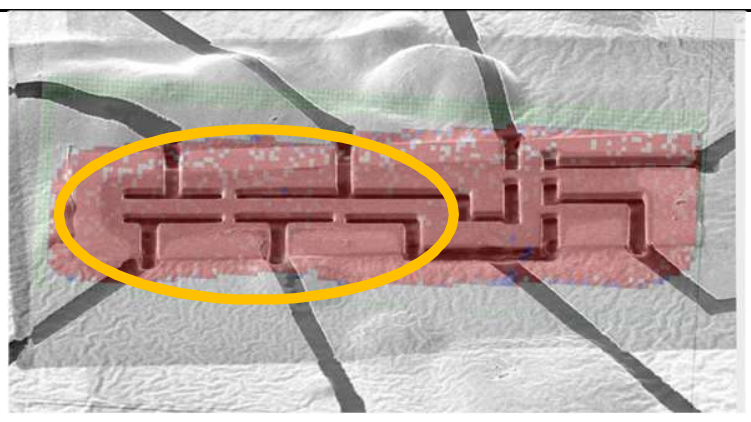 |
| $l_{xx}$ = 20.38 µm, $l_{xy}$ = 2.78 µm, $S_{xx}$ = 7.314 µm$^2$, $S_{xy}$ = 7.990 µm$^2$ | $l_{xx}$ = 10.39 µm, $l_{xy}$ = 3.22 µm, $S_{xx}$ = 4.414 µm$^2$, $S_{xy}$ = 4.594 µm$^2$ | $l_{xx}$ = 20.81 µm, $l_{xy}$ = 3.82 µm, $S_{xx}$ = 5.160 µm$^2$, $S_{xy}$ = 5.120 µm$^2$ |
| $\vec{j} \approx \parallel$[100]<br>8° – out-of-plane<br>0° – in-plane<br>([001]: 11° – out-of-plane) | $\vec{j} \approx \parallel$[100]:<br>2.5° – out-of-plane<br>10° – in-plane<br>([001]: 1° – out-of-plane) | $\vec{j} \approx \parallel$[001]<br>1° – out-of-plane<br>10° – in-plane |

* red colour – right-handed $P4_1$; blue – left handed $P4_3$ domains

**Table S3.** Measured $P4_3$-$LaRhC_2$ microdevices with the final handedness mapping (from EBSD)*, geometrical parameters of the bars [voltage separation probes length (*l*) and cross-sectional areas (*S*)], and practical orientation of the fourfold screw relative to the current flow and the lamella surface.

| **#2-1: $(\rho_{xx}, \rho_{xy}) \parallel 4_3$** | **#2-2: $(\rho_{xx}, \rho_{xy}) \parallel 4_3$** | **#3-1: $(\rho_{xx}, \rho_{xy}) \perp 4_3$** |
|---|---|---|
| 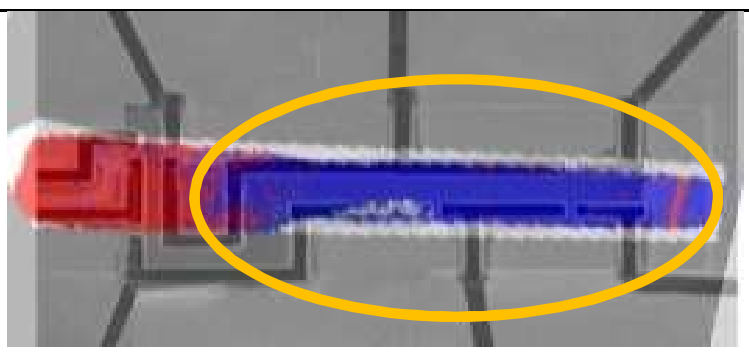 | 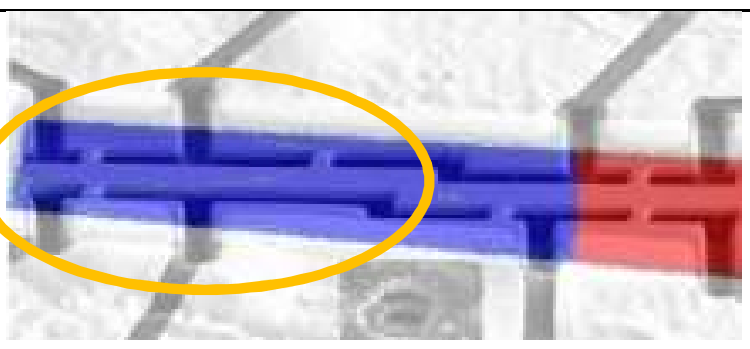 | 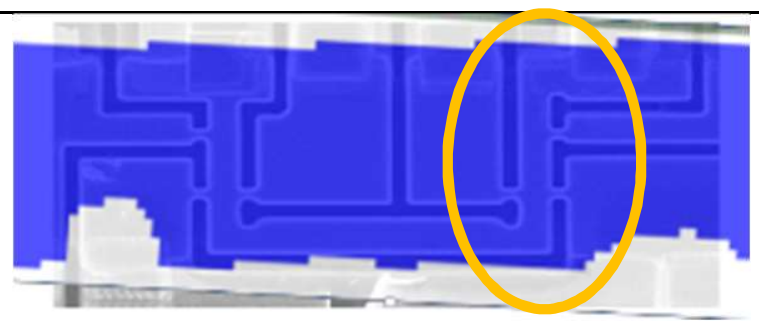 |
| $l_{xx}$ = 19.96 µm, $l_{xy}$ = 4.36 µm<br>$S_{xx}$ = 10.19 µm$^2$, $S_{xy}$ = 10.46 µm$^2$ | $l_{xx}$ = 26.06 µm, $l_{xy}$ = 4.75 µm<br>$S_{xx}$ = 7.621 µm$^2$, $S_{xy}$ = 8.120 µm$^2$ | $l_{xx}$ = 9.982 µm, $l_{xy}$ = 5.01 µm<br>$S_{xx}$ = 16.28 µm$^2$, $S_{xy}$ = 15.53 µm$^2$ |
| $\vec{j} \approx \parallel$[001]:<br>1° – out-of-plane<br>10° – in-plane | $\vec{j} \approx \parallel$[001]:<br>3° – out-of-plane<br>2° – in-plane | $\vec{j} \approx \parallel$[110]:<br>3° – out-of-plane<br>1° – in-plane<br>([001]: 3° – out-of-plane) |

* red colour – right-handed $P4_1$; blue – left handed $P4_3$ domains

**Table S4.** Details on the FIB-cut $LaRhC_2$ low-temperature single crystal data collection and analysis

| | | | | |
|---|---|---|---|---|
| Chemical formula | $LaRhC_2$ | | | |
| Formula per unit cell, *Z* | 4 | | | |
| Pearson symbol | *tP*8 | | | |
| Diffractometer | STOE STADIVARI | | | |
| Radiation and wavelength (Å) | Mo$K\alpha$; 0.71073 | | | |
| Scan type and step | Rotational method: $\omega$-scans; 1 °/frame | | | |
| Scanned $2\theta$-range (deg.) | 5.3-72.0 | | | |
| Absorption correction | Numerical | | | |
| Temperature (K) | 100(2) | | 200(2) | |
| UCPs (standardized): $a$, $c$ (Å) | 3.9650(1), 15.3206(4) | | 3.9674(1), 15.3257(4) | |
| Calculated density (g cm$^{-3}$) | 7.33(1) | | 7.32(1) | |
| Space group (No.) | $P4_1$ (76) | $P4_3$ (78) | $P4_1$ (76) | $P4_3$ (78) |
| Absorption coefficient $\mu$ (mm$^{-1}$) | 23.9 | 23.9 | 23.9 | 23.9 |
| Transmission, $T_{min}/T_{max}$ | 0.271/0.385 | 0.204/0.337 | 0.255/0.366 | 0.208/0.297 |
| No. of measured reflections | 24364 | 24611 | 24484 | 24699 |
| No. of independent reflections | 1150 | 1150 | 1150 | 1150 |
| No. of reflections with $I > 2\sigma(I)$ | 1048 | 998 | 1027 | 1012 |
| $R_\sigma$ (%) | 1.7 | 2.3 | 1.9 | 2.0 |
| $R_{int}$ (%) | 5.4 | 6.4 | 5.5 | 5.5 |
| $R_F/R_F$ for all refs. [a], (%) | 1.7/2.0 | 2.2/2.6 | 1.5/1.7 | 1.8/2.2 |
| $wR/wR$ for all refs. [b], (%) | 3.8/3.9 | 5.1/5.1 | 3.2/3.2 | 4.0/4.1 |
| Weighting scheme coefficients $m/n$ [†] | 0.021/0.229 | 0.029/0.322 | 0.014/0.024 | 0.020/0.356 |
| Extinction coefficient | 0.0129(7) | 0.0077(8) | 0.0128(6) | 0.0081(7) |
| No. of refined parameters | 27 | 27 | 27 | 27 |
| *Goodness-of-fit* | 1.12 | 1.09 | 1.18 | 1.10 |
| No. of Bijvoet pairs | 480 | 458 | 465 | 458 |
| Flack parameter | -0.01(2) | -0.03(2) | -0.03(1) | -0.01(2) |
| $\Delta\rho_{max/min}$ ($e^-$ Å$^{-3}$) | +0.93/-1.04 | +1.14/-1.14 | +0.91/-1.05 | +0.93/-1.04 |

[a] $R_F = \Sigma||F_o| - |F_c||/\Sigma|F_o|$;

[b] $wR = \{\Sigma[w(F_o^2 - F_c^2)^2/\Sigma[w(F_o^2)^2]]\}^{1/2}$;

[†] $w = 1/[\sigma^2(F_o^2) + (mP)^2 + nP]$, where $P = (F_o^2 + 2F_c^2)/3$

**Table S5.** Fractional atomic coordinates and displacement parameters of the $P4_1$-$LaRhC_2$ single crystal

| Atom | Wyckoff site | *x* | *y* | *z* | $U_{iso}/U_{eq}$ ($10^{-4}$ Å$^2$) | $U_{11}$ ($10^{-4}$ Å$^2$) | $U_{22}$ ($10^{-4}$ Å$^2$) | $U_{33}$ ($10^{-4}$ Å$^2$) | $U_{12}$ ($10^{-4}$ Å$^2$) | $U_{13}$ ($10^{-4}$ Å$^2$) | $U_{23}$ ($10^{-4}$ Å$^2$) |
|---|---|---|---|---|---|---|---|---|---|---|---|
| | | | | | ***T* = 100 K** | | | | | | |
| La | 4*a* | 0.34783(8) | 0.34820(7) | 0.0* | 31(1) | 31(1) | 30(1) | 31(2) | 0(1) | 0(1) | 0(1) |
| Rh | 4*a* | 0.8407(1) | 0.8505(1) | 0.09380(3) | 30(1) | 28(2) | 35(2) | 27(2) | 0(1) | 0(1) | 0(1) |
| C1 | 4*a* | 0.3470(14) | 0.8436(13) | 0.1370(4) | 45(9) | – | | | | | |
| C2 | 4*a* | 0.8317(13) | 0.8520(14) | 0.9590(4) | 52(9) | – | | | | | |
| | | | | | ***T* = 200 K** | | | | | | |
| La | 4*a* | 0.34743(7) | 0.34758(7) | 0.0* | 46(1) | 48(2) | 46(2) | 45(2) | 0(1) | 0(1) | 0(1) |
| Rh | 4*a* | 0.8409(1) | 0.8499(1) | 0.09381(2) | 42(1) | 39(2) | 53(2) | 34(2) | 1(1) | 2(1) | 1(1) |
| C1 | 4*a* | 0.3479(13) | 0.8455(13) | 0.1372(3) | 61(9) | – | | | | | |
| C2 | 4*a* | 0.8314(13) | 0.8513(13) | 0.9591(4) | 61(8) | – | | | | | |
| | | | | | ***T* = 293 K** | | | | | | |
| La | 4*a* | 0.34698(7) | 0.34681(7) | 0.0* | 63(1) | 64(2) | 62(2) | 64(2) | 0(1) | 0(1) | 0(1) |
| Rh | 4*a* | 0.8408(1) | 0.8492(1) | 0.09381(2) | 55(1) | 51(2) | 72(2) | 42(2) | 1(1) | 1(1) | 0(1) |
| C1 | 4*a* | 0.3480(13) | 0.8446(12) | 0.1367(3) | 68(8) | – | | | | | |
| C2 | 4*a* | 0.8340(13) | 0.8497(13) | 0.9589(4) | 81(9) | – | | | | | |

* – fixed atomic coordinate

**Table S6.** Fractional atomic coordinates and displacement parameters of the $P4_3$-$LaRhC_2$ single crystal

| Atom | Wyckoff site | *x* | *y* | *z* | $U_{iso}/U_{eq}$ ($10^{-4}$ Å$^2$) | $U_{11}$ ($10^{-4}$ Å$^2$) | $U_{22}$ ($10^{-4}$ Å$^2$) | $U_{33}$ ($10^{-4}$ Å$^2$) | $U_{12}$ ($10^{-4}$ Å$^2$) | $U_{13}$ ($10^{-4}$ Å$^2$) | $U_{23}$ ($10^{-4}$ Å$^2$) |
|---|---|---|---|---|---|---|---|---|---|---|---|
| | | | | | ***T* = 100 K** | | | | | | |
| La | 4*a* | 0.6521(1) | 0.6522(1) | 0.0* | 37(1) | 39(2) | 37(2) | 36(2) | 0(1) | 1(1) | 1(1) |
| Rh | 4*a* | 0.1585(1) | 0.1496(2) | 0.90622(4) | 35(1) | 33(3) | 39(3) | 32(3) | 0(2) | 2(2) | 1(2) |
| C1 | 4*a* | 0.1649(19) | 0.1488(19) | 0.0406(5) | 57(12) | | | – | | | |
| C2 | 4*a* | 0.651(2) | 0.1538(19) | 0.8631(5) | 69(13) | | | – | | | |
| | | | | | ***T* = 200 K** | | | | | | |
| La | 4*a* | 0.65245(9) | 0.65283(9) | 0.0* | 51(1) | 53(2) | 50(2) | 49(2) | 1(1) | 1(1) | 2(1) |
| Rh | 4*a* | 0.1584(1) | 0.1502(1) | 0.90621(3) | 45(1) | 40(2) | 53(2) | 43(2) | 0(1) | 2(2) | 1(2) |
| C1 | 4*a* | 0.1669(16) | 0.1493(16) | 0.0406(4) | 63(10) | | | – | | | |
| C2 | 4*a* | 0.6530(17) | 0.1540(16) | 0.8626(4) | 81(11) | | | – | | | |
| | | | | | ***T* = 293 K** | | | | | | |
| La | 4*a* | 0.65285(9) | 0.65349(9) | 0.0* | 60(1) | 61(2) | 59(2) | 60(2) | 1(1) | 1(1) | 0(1) |
| Rh | 4*a* | 0.1586(1) | 0.1508(1) | 0.90621(3) | 51(1) | 44(2) | 63(2) | 47(2) | 1(1) | 2(1) | 0(1) |
| C1 | 4*a* | 0.1674(16) | 0.1485(16) | 0.0406(4) | 66(9) | | | – | | | |
| C2 | 4*a* | 0.6538(16) | 0.1551(15) | 0.8626(4) | 83(10) | | | – | | | |

* – fixed atomic coordinate

**Table S7.** Interatomic distances in the studied $LaRhC_2$ single crystals at different temperatures

| **Temperature** | **100 K** | | **200 K** | | **293 K** | |
|---|---|---|---|---|---|---|
| **Space group** | $P4_1$ | $P4_3$ | $P4_1$ | $P4_3$ | $P4_1$ | $P4_3$ |
| **Atoms** | **Distances (Å)** | | **Distances (Å)** | | **Distances (Å)** | |
| La–La | 4.0153(2) | 4.0156(2) | 4.0180(2) | 4.0183(2) | 4.0213(2) | 4.0214(2) |
| La–La | 3.9650(1) | 3.9650(1) | 3.9674(1) | 3.9674(1) | 3.9703(1) | 3.9703(1) |
| La–Rh | 3.1740(5) | 3.1730(7) | 3.1746(5) | 3.1737(6) | 3.1758(5) | 3.1754(6) |
| La–Rh | 3.1625(5) | 3.1602(8) | 3.1629(5) | 3.1607(7) | 3.1640(5) | 3.1621(6) |
| La–Rh | 3.1387(5) | 3.1406(8) | 3.1419(5) | 3.1439(7) | 3.1450(5) | 3.1466(6) |
| La–Rh | 3.1271(5) | 3.1278(7) | 3.1301(5) | 3.1308(6) | 3.1332(5) | 3.1332(6) |
| La–C | 2.928(6) | 2.917(8) | 2.929(5) | 2.923(6) | 2.921(5) | 2.929(6) |
| La–C | 2.908(5) | 2.900(8) | 2.909(5) | 2.904(6) | 2.909(5) | 2.906(6) |
| La–C | 2.907(5) | 2.898(7) | 2.901(5) | 2.900(7) | 2.905(5) | 2.902(6) |
| La–C | 2.900(5) | 2.891(8) | 2.896(5) | 2.896(6) | 2.896(5) | 2.899(6) |
| La–C | 2.885(5) | 2.890(8) | 2.890(5) | 2.892(7) | 2.893(5) | 2.894(6) |
| La–C | 2.875(5) | 2.882(8) | 2.885(5) | 2.890(7) | 2.881(5) | 2.890(6) |
| La–C | 2.840(5) | 2.847(7) | 2.841(5) | 2.844(6) | 2.850(5) | 2.850(6) |
| La–C | 2.819(6) | 2.828(8) | 2.821(6) | 2.824(6) | 2.834(5) | 2.822(6) |
| Rh–C | 2.146(6) | 2.150(8) | 2.149(5) | 2.153(6) | 2.151(6) | 2.152(6) |
| Rh–C | 2.114(6) | 2.116(8) | 2.118(5) | 2.114(7) | 2.119(5) | 2.113(6) |
| Rh–C | 2.067(6) | 2.063(8) | 2.066(5) | 2.073(7) | 2.069(5) | 2.077(6) |
| Rh–C | 2.066(6) | 2.060(8) | 2.065(5) | 2.059(7) | 2.064(5) | 2.061(6) |
| C–C | 1.357(8) | 1.364(11) | 1.357(7) | 1.356(9) | 1.358(7) | 1.356(9) |

**Table S8.** Standardized UCPs of the polycrystalline $LaRhC_2$ at different temperatures (synchrotron powder diffraction)

| $T$ (K) | $a$ (Å) | $c$ (Å) | $V$ (Å$^3$) | $T$ (K) | $a$ (Å) | $c$ (Å) | $V$ (Å$^3$) |
|---|---|---|---|---|---|---|---|
| 295.1(4) | 3.9704(1) | 15.3323(3) | 241.69(1) | 192.0(4) | 3.9672(1) | 15.3252(3) | 241.20(1) |
| 292.1(4) | 3.9703(1) | 15.3322(3) | 241.68(1) | 189.0(4) | 3.9671(1) | 15.3250(3) | 241.18(1) |
| 288.9(4) | 3.9702(1) | 15.3320(3) | 241.66(1) | 185.8(4) | 3.9670(1) | 15.3249(3) | 241.17(1) |
| 286.0(4) | 3.9701(1) | 15.3317(3) | 241.65(1) | 182.9(4) | 3.9670(1) | 15.3247(3) | 241.16(1) |
| 283.0(4) | 3.9700(1) | 15.3315(3) | 241.63(1) | 179.9(4) | 3.9669(1) | 15.3245(3) | 241.15(1) |
| 279.9(4) | 3.9699(1) | 15.3312(3) | 241.62(1) | 176.9(4) | 3.9668(1) | 15.3244(3) | 241.13(1) |
| 277.0(5) | 3.9698(1) | 15.3310(3) | 241.60(1) | 173.8(4) | 3.9667(1) | 15.3242(3) | 241.12(1) |
| 273.9(4) | 3.9697(1) | 15.3308(3) | 241.58(1) | 170.8(4) | 3.9667(1) | 15.3240(3) | 241.11(1) |
| 270.9(4) | 3.9696(1) | 15.3305(3) | 241.57(1) | 167.7(5) | 3.9666(1) | 15.3239(3) | 241.10(1) |
| 267.7(5) | 3.9695(1) | 15.3303(3) | 241.55(1) | 164.8(4) | 3.9665(1) | 15.3237(3) | 241.09(1) |
| 264.8(4) | 3.9694(1) | 15.3301(3) | 241.54(1) | 161.6(4) | 3.9664(1) | 15.3236(3) | 241.07(1) |
| 261.7(4) | 3.9693(1) | 15.3299(3) | 241.52(1) | 158.7(5) | 3.9664(1) | 15.3234(3) | 241.06(1) |
| 258.7(4) | 3.9692(1) | 15.3296(3) | 241.51(1) | 155.7(4) | 3.9663(1) | 15.3233(3) | 241.05(1) |
| 255.7(5) | 3.9691(1) | 15.3294(3) | 241.49(1) | 152.7(5) | 3.9662(1) | 15.3231(3) | 241.04(1) |
| 252.7(5) | 3.9690(1) | 15.3292(3) | 241.48(1) | 149.6(4) | 3.9661(1) | 15.3230(3) | 241.03(1) |
| 249.8(5) | 3.9689(1) | 15.3290(3) | 241.46(1) | 146.7(5) | 3.9661(1) | 15.3228(3) | 241.02(1) |
| 246.5(4) | 3.9688(1) | 15.3288(3) | 241.45(1) | 143.6(5) | 3.9660(1) | 15.3226(3) | 241.00(1) |
| 243.6(5) | 3.9687(1) | 15.3286(3) | 241.43(1) | 140.5(4) | 3.9659(1) | 15.3224(3) | 240.99(1) |
| 240.4(4) | 3.9686(1) | 15.3284(3) | 241.42(1) | 137.5(4) | 3.9658(1) | 15.3223(3) | 240.98(1) |
| 237.6(5) | 3.9685(1) | 15.3282(3) | 241.40(1) | 134.4(4) | 3.9658(1) | 15.3221(3) | 240.97(1) |
| 234.5(4) | 3.9685(1) | 15.3280(3) | 241.39(1) | 131.5(4) | 3.9657(1) | 15.3220(3) | 240.96(1) |
| 231.5(4) | 3.9684(1) | 15.3278(3) | 241.38(1) | 128.4(4) | 3.9656(1) | 15.3219(3) | 240.95(1) |
| 228.4(4) | 3.9683(1) | 15.3275(3) | 241.36(1) | 125.4(4) | 3.9656(1) | 15.3218(3) | 240.94(1) |
| 225.4(4) | 3.9682(1) | 15.3273(3) | 241.35(1) | 122.5(5) | 3.9655(1) | 15.3216(3) | 240.93(1) |
| 222.4(4) | 3.9681(1) | 15.3271(3) | 241.33(1) | 119.9(5) | 3.9654(1) | 15.3215(3) | 240.92(1) |
| 219.3(4) | 3.9680(1) | 15.3269(3) | 241.32(1) | 117.6(5) | 3.9654(1) | 15.3214(3) | 240.91(1) |
| 216.4(5) | 3.9679(1) | 15.3267(3) | 241.30(1) | 115.5(4) | 3.9653(1) | 15.3213(3) | 240.91(1) |
| 213.2(4) | 3.9678(1) | 15.3266(3) | 241.29(1) | 113.4(4) | 3.9653(1) | 15.3212(3) | 240.90(1) |
| 210.2(4) | 3.9677(1) | 15.3264(3) | 241.28(1) | 111.4(5) | 3.9653(1) | 15.3211(3) | 240.89(1) |
| 207.2(4) | 3.9676(1) | 15.3262(3) | 241.26(1) | 109.4(4) | 3.9652(1) | 15.3211(3) | 240.89(1) |
| 204.1(4) | 3.9676(1) | 15.3259(3) | 241.25(1) | 107.6(4) | 3.9652(1) | 15.3210(3) | 240.88(1) |
| 201.1(4) | 3.9675(1) | 15.3258(3) | 241.23(1) | 105.9(4) | 3.9651(1) | 15.3209(3) | 240.88(1) |
| 198.0(4) | 3.9674(1) | 15.3256(3) | 241.22(1) | 104.3(5) | 3.9651(1) | 15.3209(3) | 240.87(1) |
| 195.0(4) | 3.9673(1) | 15.3254(3) | 241.21(1) | 103.0(5) | 3.9651(1) | 15.3208(3) | 240.87(1) |
| Extrapolated: | | | | 100.0 | 3.9650(1) | 15.3206(4) | 240.86(1) |
| | | | | 0 | 3.9632(1) | 15.3170(4) | 240.58(1) |

**Table S9.** Selected temperature ranges and estimated from Arrhenius fits* energy gaps for the $P4_1$- and $P4_3$-$LaRhC_2$ microdevices.

| Microdevice number | #1-2 | #2-1 | #2-2 | #1-1 | #1-2 | #3-1 |
|---|---|---|---|---|---|---|
| Orientation† $\rho_{xx}$ with respect to the 4-fold screw axis | $\parallel 4_1$ $\vec{j} \parallel [001]$ | $\parallel 4_3$ $\vec{j} \parallel [001]$ | $\parallel 4_3$ $\vec{j} \parallel [001]$ | $\perp 4_1$ $\vec{j} \parallel [100]$ | $\perp 4_1$ $\vec{j} \parallel [100]$ | $\perp 4_3$ $\vec{j} \parallel [110]$ |
| Temperature range (K) | | | 180-300 | | | |
| $\rho_0$ (mΩ cm) | 2.24(2) | 2.69(3) | 2.89(2) | 0.82(2) | 0.92(2) | 1.14(3) |
| $E_g^{\rho}$ (meV) | 21(1) | 20(1) | 19(1) | 32(2) | 31(2) | 35(2) |

* $\rho_{xx}(T) = \rho_0 exp(\frac{E_g^{\rho}}{2k_B T})$

† Specifically, Tables S2,S3